\documentclass[12pt,a4paper]{article}

\usepackage{lmodern}[lmr]
\usepackage{appendix}
\usepackage{graphicx}
\usepackage{arydshln}
\usepackage{pdflscape}
\usepackage{caption}
\usepackage{bm,amsfonts, amsmath, amssymb, amsthm, color, enumitem, float, mathrsfs, mathtools, multirow, nccmath, rotating, tikz, tocloft, setspace}
\usepackage{etoolbox}
\usepackage{subcaption}
\usepackage[mathcal]{euscript}
\usepackage[colorlinks=true,allcolors=blue]{hyperref}
\usepackage[nameinlink,noabbrev,capitalize]{cleveref}
\usepackage[hmargin=1in, vmargin=0.9in]{geometry}
\usepackage{threeparttable}

\usepackage[round]{natbib}
\def\be{\begin{equation}}
\def\ee{\end{equation}}
\def\bea{\begin{eqnarray}}
\def\eea{\end{eqnarray}}

\author{}
\title{}

\newtheorem{assumption}{Assumption}

\newtheorem{lemma}{Lemma}

\newtheorem{theorem}{Theorem}
\graphicspath{ {./files/}} 

\numberwithin{equation}{section}
\numberwithin{lemma}{section}

\newcommand\blfootnote[1]{
\begingroup
\renewcommand\thefootnote{}\footnote{#1}
\addtocounter{footnote}{-1}
\endgroup
}

\allowdisplaybreaks[4]

\begin{document}

\setlength{\abovedisplayskip}{8pt}
\setlength{\belowdisplayskip}{8pt}

\begin{titlepage}

\begin{center}

\begin{spacing}{1.5}
{\Large  \textbf{Binary Response Forecasting under a Factor-Augmented Framework}}

\end{spacing}

\bigskip

$^{\ast}${\sc Tingting Cheng}, $^{\dag}${\sc Jiachen Cong}, $^{\ast}${\sc Fei Liu} and $^{\ast}${\sc Xuanbin Yang}\blfootnote{Correspondence to: Fei Liu, School of Finance,
Nankai University, Tianjin, China; email: feiliu.econ@outlook.com. Cheng acknowledges the financial support from the National Natural Science Foundation of China (72173068, 71803091), and the Fundamental Research Funds for the Central Universities. Liu's research is financially supported by the National Natural Science Foundation of China under the grant number: 72203114. }   

\bigskip

$^{\ast}$School of Finance, Nankai University, China

\medskip

$^{\dag}$Department of Statistics, University of Illinois Urbana-Champaign, United States

\bigskip

\today

\end{center}

\begin{abstract}
 
In this paper, we propose a novel factor-augmented forecasting regression model with a binary response variable. We develop a maximum likelihood estimation method for the regression parameters and establish the asymptotic properties of the resulting estimators. Monte Carlo simulation results show that the proposed estimation method performs very well in finite samples. Finally, we demonstrate the usefulness of the proposed model through an application to  U.S. recession forecasting. The proposed model consistently outperforms conventional Probit regression across both in-sample and out-of-sample exercises, by effectively utilizing high-dimensional information through latent factors.

\medskip
	

\medskip
	
	
\end{abstract}

\end{titlepage}

\section{Introduction}\label{section-intro}

In recent years, factor-augmented forecasting regression (FAR) models have received much attention in economic and finance following the seminal work of \cite{Stock01122002}. They show that principal components extracted from a large set of predictors can substantially enhance forecasting accuracy of economic variables. Building on this, \cite{baing2006} establish the theory for the estimated parameters of the factor-augmented regressions to enable statistical inference. Following their work, researchers have conducted many theoretial and applied work for FAR. For instance, on the theoretical side, \cite{BaiAndNg2008} develop variable selection methods using thresholding rules to refine factor construction; \cite{XuCheng2015} propose a forecast combination approach within the FAR framework, assigning weights to individual models based on the Mallows and leave-h-out cross-validation criteria. Recently, 
\cite{Hannadige2024} propose the mixture-FAR method to incorporate both stationary and nonstationary factors. \cite{fanjianqing2024} propose FAR-NN model, which uses a diversified projection matrix to linearly compress observed variables to get a surrogate for latent factors and fits the regression function via deep ReLU networks. 
\cite*{chenfanzhu2024} extend FAR from vector covariates to matrix covariates and introduce a matrix factor model structure. Other theoretical advances were made, among others, by \cite{wangcuili2015}, \cite{PerronMcCracken}, \cite{Karabiyik2018}, \cite{Djogbenou2020} and \cite{TuWang2025modelselection}. On the empirical side, \cite{Dijk2016} use factors extracted from a large set of macroeconomic variables to predict monthly U.S. excess stock returns. \cite{KellySeth2015} show that the factor-augmented
 approach can improve the conventional predictive regression forecasts in out-of-sample tests. We can also see the usefulness of factor-augmented models for forecasting finance and economics data from the discussions in \cite{Aastveit2015}, \cite{Swanson2020} and \cite{Qiu2020}.

It should be noticed that the existing statistical theory has primarily focused on factor-augmented forecasting regression with continuous response variables \citep[e.g.][]{baing2006,XuCheng2015,Hannadige2024}. However, an increasing number of studies have shown that binary response variable models have been widely applied in the fields of economics and finance, such as the prediction of bank failure \citep[e.g.][]{audrino2018predicting, kim2022predicting}, corporate bankruptcy, default, and financial distress \citep[e.g.][]{li2020predicting,Charalambous2023},  credit default \citep[e.g.][]{liuzhangxiong2024} and business circle\citep[e.g.,][]{bybee2024business}.

Despite the large number of potential applications, to the best of our knowledge the theoretical framework of the factor-augmented model with binary response variables is undeveloped. In this paper, we aim to fill this gap in the literature. Specifically, we consider the model:\footnote{Recent studies have also explored the incorporation of non-linear features and time-varying coefficients into factor-augmented regression models. For example, \citet{yancheng2022} introduce threshold effects into the factor-augmented framework; \citet{massacci2024} examine the forecasting performance of such models in the presence of structural breaks; \citet{Lietal2020} propose a nonlinear functional-coefficient model where regression coefficients vary smoothly with an index variable; and \citet*{chenhong2024} develop a time-varying forecast combination approach for factor-augmented regressions with smooth structural changes. In this paper, we focus on addressing the challenge of binary response variables within the factor-augmented framework, and we leave the exploration of more flexible model structures to future research.}
    \begin{align}\label{introduction_1}
    y_{t+h}=
    \begin{cases}
        1 , \  \mbox{if} \ \beta_0+\beta_w^\top w_t+\beta_f^\top f_t-\epsilon_{t+h}\geq0; \\
        0, \  \mbox{otherwise},
    \end{cases}
\end{align}
where $t=1,2,\cdots,T-h$, $h\geq0$ is the forecast horizon, $y_{t+h}$ is the binary dependent variable, $w_{t}$ is a $p\times1$ vector of observed regressors, $f_{t}$ is a $d_0\times1$ vector of unobserved  factors, $(\beta_0,\beta_w^\top,\beta_f^\top)^\top$ are unknown parameters to be estimated, and $\epsilon_{t+h}$ is the idiosyncratic error term. In addition, we assume that both $p$ and $d_0$ are known and define $k=p+d_0+1$ as the total number of observable and unobservable factors. 
Suppose that we observe a large number of predictors $x_{it}$ that contain information about $f_{t}$, which is captured by a factor model as follows:
\begin{align}\label{introduction_2}
    x_{it}=\lambda_{i}^{\top}f_{t}+e_{it}, \  \ \text{for~} i=1,2,\cdots,N, \text{and} \ t=1,2,\cdots, T,
\end{align}
where $\lambda_{i}$ is the factor loadings and $e_{it}$ is another idiosyncratic error.  

From a modelling standpoint, our framework is closely related to those of \citet{baing2006} and \cite{Hannadige2024}, but with a key distinction that we consider a binary response variable.  Unlike the continuous‐response framework, the binary nature of our dependent variable induces two fundamental deficiencies in their least‐squares approaches: (i) fitted probabilities can fall outside the unit interval, and (ii) the regression errors are inherently heteroskedastic.  To remedy these shortcomings, we adopt a maximum‐likelihood framework and establish the resulting estimators’ consistency and asymptotic normality.  Moreover, we relax some of the original assumptions by introducing  $\alpha$-mixing conditions to capture the temporal dependence in the variables. 
Our model also shares structural similarities with that of \cite{gao2023}. Both studies consider the binary response variable and MLE method for estimation. However, \cite{gao2023} focus on a panel data model in which the factors and model parameters are jointly estimated by MLE. Their specification relies on a fixed, low‐dimensional set of regressors and thus cannot accommodate a large number of covariates. In contrast, our approach applies principal component analysis (PCA) to a growing number of observed predictors, thereby enabling consistent forecasting in high‐dimensional settings.


From a practical perspective, our model is well-suited for classification tasks commonly encountered in economics and finance. Typical examples include applying high-dimensional features or large-scale factor structures for binary outcome prediction tasks, such as corporate bankruptcy forecasting in financial markets \citep[e.g.][]{Jones2017highdimensional,JonesStewartJohnstone2017}. Additionally, the model can effectively incorporate a wide range of high-dimensional variables---such as financial ratios, macroeconomic indicators, and transactional data---for credit risk assessment and credit scoring \citep[e.g.][]{Laitinen1999,Masten2021,Silva2022,Liuyanghuangmazheng2024}.
In our empirical analysis, we focus on forecasting U.S. recessions. Using a large monthly macroeconomic dataset spanning from January 1960 to December 2024, we show that our proposed model consistently outperforms conventional Probit regression across both in-sample and out-of-sample exercises. In particular, our model delivers superior predictive accuracy at short- and medium-term horizons, effectively identifying major recessions such as the 2008 subprime crisis and the 2020 COVID-19 pandemic. These results demonstrate the practical relevance and applicability of the proposed model and estimation method, highlighting the value of incorporating high-dimensional information through latent factors in binary prediction tasks involving macroeconomic turning points.

In summary, the main contributions of the paper are as follows. (i) We develop a binary response factor-augmented forecasting regression that extends the original FAR framework \citep[e.g.,][]{baing2006,Hannadige2024} to accommodate binary response variables, thereby broadening its applicability to classification-oriented forecasting problems. (ii) We propose a maximum likelihood estimation method to estimate the unknown parameters and establish the corresponding asymptotic properties. In addition, different from \cite{baing2006}, our theoretical results are derived under $\alpha$-mixing condition of $\{w_t, f_t, x_{it},\epsilon_{t},e_{it}\}$, which represents a more general dependence structure \citep[e.g.,][]{FanYao,Gao2007}. (iii) We examine the finite-sample performance of the estimators through simulation studies and further use the proposed model to forecast U.S. recessions,  highlighting its real-world relevance.

The structure of this paper is as follows. Section \hyperref[section-methods]{2} introduces the proposed model, develops the estimation methodology, and establishes
 the asymptotic theory. Section \hyperref[simulation]{3} conducts simulation experiments to examine the finite-sample performance of the
 theoretical findings. Section \hyperref[a_case_study]{4} provides an empirical study to demonstrate the model's practical utility by applying it to forecast U.S. recessions. Section \hyperref[conclusion]{5} concludes. The proof of the main theorems is provided in the appendix.

 Before proceeding further, we introduce some mathematical symbols that will be used repeatedly throughout the paper. $\| \cdot \|$ denotes the Euclidean norm of a vector or the Frobenius norm of a matrix; for a vector or a matrix $A$, $A^\top$ denotes the transpose of $A$; $\mathrm{diag}(A_1, \ldots, A_n)$ represents a block-diagonal matrix with its diagonal elements are $A_1,\cdots, A_n$, and $I_m$ denotes an identity matrix of dimension $m$. In addition, $\xrightarrow{P}$ and $\xrightarrow{D}$ denote convergence in probability and in distribution, respectively. For a square matrix $M$, $\rho_{\min}(M)$ stands for its smallest eigenvalue; $a \wedge b = \min\{a, b\}$ and $a \vee b = \max\{a, b\}$.

 \section{Estimation and asymptotic theory}\label{section-methods}

In this section, we introduce our proposed estimation method and then establish the corresponding asymptotic theory.
 
 \subsection{Model and estimation}\label{section-est}
 
 Before proceeding further,  we first introduce some additional notations. Let  $\beta=(\beta_0,\beta_w^\top,\beta_f^\top)^\top$, $z_t=(1, w_t^\top, f_t^\top)^\top$, $X=(X_{1},\ldots, X_{N})^\top$ with $X_i=(x_{i1},\ldots, x_{iT})^\top$, $F = (f_1,\ldots, f_T)^\top$ is a $T\times d_0$ matrix consisting of factor regressors, $\Lambda=(\lambda_1,\lambda_2,\cdots, \lambda_N)^\top$ is a $N\times d_0$ matrix and $E$ denotes a $T\times N$ error matrix containing the elements $\{e_{it}\}$. With these notations, Model (\ref{introduction_1}) and (\ref{introduction_2})  can be rewritten as 
     \begin{align*}
    &y_{t+h}=
    \begin{cases}
        1 , \  \mbox{if} \ \beta^{\top}z_{t}-\epsilon_{t+h}\geq0 \\
        0, \  \mbox{otherwise}
    \end{cases}\\
    &X=F\Lambda^\top+E.
\end{align*}
 Noteworthily, we assume the true number of factors $d_0$ is known in this step of estimation and will formally discuss the factor number selection issue in Section \ref{section-facno}.

  In this paper, our object of interest is to estimate the unknown parameter vector $\beta$ in model (\ref{introduction_1}) and the conditional probability $\mathbb{P}(y_{t+h}=y \vert z_{t})$. To this end, we introduce a three-step estimation procedure, which is given as follows.
 
 \medskip 
 {\bf Step 1}: PCA estimation of unknown factors. 
 We follow the methodology of  \cite{bai2003} and apply principal component analysis to estimate the unknown factors. Let $\tilde{F}=(\tilde f_1,\ldots, \tilde f_T)^\top$ denote the estimated factor matrix. Specifically,
 \begin{align*}
     \frac{1}{NT}XX^\top\tilde{F}=\tilde{F}V_{NT},
 \end{align*}
 where $V_{NT}$ is a $d_0\times d_0$ diagonal matrix that contains the first $d_0$ largest eigenvalues of $\frac{1}{NT}XX^{\top}$ in descending order, $\Tilde{F}$ is a $T\times d_0$ matrix that contains the corresponding eigenvectors and satisfies $ \frac{1}{T}\tilde{F}^\top\tilde{F}=I_{d_0}$.

{\bf Step 2}: MLE estimation of $\beta$.  Simple algebra yields     
\begin{align*}
    \mathbb{P}(y_{t+h}=y \vert z_{t})=\Phi_{\epsilon}(\beta^{\top}z_{t})^{y}[1-\Phi_{\epsilon}(\beta^{\top}z_{t})]^{1-y}.
\end{align*}
    for $y=0$ or $1$,   where $\Phi_{\epsilon}(\cdot)$ is the CDF of $\epsilon_{t}$. 
  Drawing upon this fact, by replacing the unknown factors with their estimators, we obtain the factor augmented log-likelihood function:
\begin{align*}
        \log[L(\beta)]=\sum_{t=1}^{T-h}[y_{t+h} \log\Phi_{\epsilon}(\beta^{\top}\tilde{z_{t}})+(1-y_{t+h}) \log(1-\Phi_{\epsilon}(\beta^{\top}\tilde{z_{t}}))],
\end{align*}
where $\tilde{z_{t}}=(1,w_{t}^{\top},\tilde{f}_{t}^{\top})^{\top}$.
 
    Therefore, we can construct the maximum likelihood estimator of $\beta$:
    \begin{align*}
        \widehat{\beta}=\arg\max\limits_{\beta}\log[L(\beta)],
    \end{align*}
    where we define $\widehat{\beta}=(\widehat{\beta}_0,\widehat{\beta}_{w}^{\top},\widehat{\beta}_f^{\top})^{\top}$.

{\bf Step 3}: Construction of predicted probabilities. Given the PCA estimates  \(\tilde{F}\) and the MLE \(\widehat{\beta}\), we compute the predicted probability of the binary outcome:
\[
    \widehat{\mathbb{P}}(y_{t+h} = 1 \mid \tilde{z}_{t}) 
    = \Phi_{\epsilon}(\widehat{\beta}^\top \tilde{z}_{t}).
\]

\medskip 
In the first step of our procedure, we extract factor estimates from a large set of continuous predictor variables via the PCA method.  Alternatively, if some of the predictors in $x_{it}$ are discrete, we can apply the MLE approach of \cite{wang2020} for nonlinear factor models to construct factor proxies.  Specifically, we may consider the following nonlinear factor structure:
\[
\mathbb{P}(x_{it}=x|\lambda_i^\top f_t)=g_{it}(x,\lambda_i^\top f_t),
\]
where \(g_{it}(x,\cdot)\) is some known joint probability mass function of $x_{it}$ given the factors and loadings. The factor estimators can then be obtained by maximizing the following log-likelihood function:
\[
    (\tilde{\Lambda}, \tilde{F})=\arg\max_{\Lambda,F}\sum_{i=1}^N \sum_{t=1}^T\log g_{it}(x_{it}, \lambda_i^\top f_t),
\]
Under suitable regularity conditions, \cite{wang2020} show that these maximum likelihood estimators achieve asymptotic properties comparable to those of the PCA estimators in linear factor models. In this regard, our proposed method can be extended to the case with discrete predictor variables by replacing the PCA step in the first stage with this nonlinear MLE-based factor estimation approach.

In the next subsection, we establish the asymptotic theory for $\widehat{\beta}$.

\subsection{Asymptotic theory}\label{section-asy}
\noindent

To facilitate the theoretical developments, we further define some notations. Let  $E_t = (e_{1t},\ldots, e_{Nt})^\top$, $H=(\Lambda^\top\Lambda/N)(F^\top\tilde{F}/T)V_{NT}^{-1}$, $f_t^\circ=H^\top f_t$, $z^{\circ}_{t}=(1,  w_{t}^\top, (H^\top f_{t})^\top)^\top$ and  $\beta^{\circ}=(\beta_0,\beta_w^{\top},\beta_f^\top(H^\top)^{-1})^{\top}$. It is straightforward to see that $z_{t}^{\circ}=\text{diag}(I_{p+1},H^\top)z_t$,  $\beta^\circ=\text{diag}(I_{p+1},H^{-1})\beta$ and $ \beta^\top z_t=\beta^{\circ\top} z_t^{\circ}$. Additionally, $M$ represents a sufficiently large constant that could be different at each appearance.

\begin{assumption}\label{assumption_1}~
\begin{enumerate}
    \item  $\{w_t,f_{t},\epsilon_{t}\}$ are strictly stationary and $\alpha$-mixing with the $\alpha$-mixing coefficient $\alpha(|t-s|)$  satisfying that $\sum_{t=0}^\infty(\alpha(t))^{\delta/(4+\delta)}=O(1)$, where $\delta$ satisfies $\mathbb{E}\|f_t\|^{4+\delta}\leq M$.

    \item The loadings $\{\lambda_i\}$ satisfy that $\mathbb{E}\|\lambda_i\|^{4+\delta} \leq M$. In addition,  $\frac{1}{N}\Lambda^\top\Lambda=\Sigma_\lambda+O_P\big(\frac{1}{\sqrt{N}}\big)$, where $\Sigma_\lambda$ is a positive definite matrix.
    \item  $\{e_{it}\}$ are strictly stationary and $\alpha$-mixing across $t$ with the $\alpha$-mixing coefficient $\alpha_{ij}(|t-s|)$ satisfying that $\sum_{i,j=1}^N\sum_{t=0}^\infty(\alpha_{ij}(t))^{\delta/(4+\delta)}=O(N)$ and $\max_{i \ge 1} \sum_{t=1}^{\infty} (\alpha_{ii}(t))^{\delta/(4+\delta)}$ $ = O(1)$. Additionally,  $\mathbb{E}|e_{it}|^{4+\delta}\leq M$.

\item $\{w_t,\lambda_i,f_t\}$,  $\{e_{it}\}$ and $\{\epsilon_{t}\}$ are mutually independent.
\end{enumerate}
\end{assumption}

\hyperref[assumption_1]{Assumption 1} imposes $\alpha$-mixing conditions to control the serial and cross-sectional dependence among the variables. Such conditions are standard in the literature on factor models (see, for example, Assumption 3.1 in \cite{gao2023} and Assumption 1.(i) of \cite{GAO2020329}). In addition, we assume strict exogeneity by requiring that the variables $w_t, \lambda_i$, and $f_t$ are independent of the random errors. It is worth noting that dependence between the observable and unobservable predictors is permitted.


With \hyperref[assumption_1]{Assumption 1}, we establish the consistency of the factor estimators obtained in the first step of our estimation procedure.

\begin{lemma}\label{lemma_2.2}  Let \hyperref[assumption_1]{Assumption 1} hold. As $N,T\rightarrow\infty$,
     \begin{equation*}
         \frac{1}{T} \sum_{t=1}^{T} \left\| \tilde{z}_t - z_t^\circ \right\|^2 = O_P\left(\frac{1}{N \wedge T}\right).
     \end{equation*}
\end{lemma}

Lemma \ref{lemma_2.2} follows directly from $\frac{1}{T} \sum_{t=1}^{T} \left\| \tilde{f}_t - H^\top f_t \right\|^2=O_P\left(\frac{1}{N\wedge T}\right)$, which characterizes the rate of convergence for the factor estimators obtained via the PCA method. This rate is consistent with the seminal result of \cite{bai2003} and extends their framework to settings with $\alpha$-mixing dependence. Notably, Lemma \ref{lemma_2.2}  provides the theoretical foundation for our augmented estimation of the coefficient  $\beta$ in the second step of our procedure.

For notational convenience, let \( u_t = \beta^\top z_t \) and \( \widehat{u}_t = \widehat{\beta}^\top \tilde{z}_t \). Additionally, define the log-likelihood contribution at time \( t \) as
$l_t(x) = y_{t+h} \log \Phi_\epsilon(x) + (1 - y_{t+h}) \log (1 - \Phi_\epsilon(x))$,
where \( \Phi_\epsilon(\cdot) \) is the distribution function of the error term \( \epsilon \). Let \( l_t'(x) \), \( l_t''(x) \), and \( l_t'''(x) \) denote the first, second, and third derivatives of \( l_t(x) \) with respect to \( x \), respectively.

\begin{assumption}\label{assumption_2}~
\begin{enumerate}
    \item Suppose that $\{\epsilon_{it}\}$ is identically distributed with known PDF and CDF as $\phi_{\epsilon}(\cdot)$ and $\Phi_{\epsilon}(\cdot)$, respectively.
    \item There exists a set $\Xi_{T}=[\Xi_{T}^l, \Xi_{T}^u]$ and  a positive constant $c$ such that all $\beta^\top z_t$'s belong to $\Xi_{T}$ with probability approaching 1,  $\inf_{z\in\Xi_T} |\phi_{\epsilon}(z)|>c$ and $0<\Phi_\epsilon(\Xi_{T}^l)<\Phi_\epsilon(\Xi_{T}^u)<1$.  
    \item  \(\phi_{\epsilon}(\cdot)\) is twice continuously differentiable on \(\Xi_{T}\), with both its first and second derivatives uniformly bounded. 
    \item  \(l_{t}(x)\) is three times continuously differentiable, with uniformly bounded derivatives up to third order. Additionally, there exists $b_U > b_L > 0$ such that $b_L \leq -l_t''(x)  \leq b_U$.
    \item There exists a sufficiently large positive number $M$ such that $\Vert \beta \Vert \leq M$.
\end{enumerate}
\end{assumption}

\hyperref[assumption_2]{Assumption 2} imposes standard conditions for maximum likelihood estimation. Specifically, \hyperref[assumption_2]{Assumption 2.1} requires that the innovation terms are identically distributed with a known distribution.  This assumption can be relaxed by allowing the error distribution to be unknown, in which case its PDF and CDF can be estimated nonparametrically.  \hyperref[assumption_2]{Assumption 2.2} imposes uniform bounds on the support of \(\beta^\top z_t\) with probability approaching one.  Assuming bounded support is conventional in the MLE literature for nonlinear models \citep[e.g.,][]{wang2020}.  Additionally, \hyperref[assumption_2]{Assumption 2.3} and \hyperref[assumption_2]{2.4} regulate the derivatives of the error PDF and the individual log‐likelihood function, ensuring the existence and nonsingularity of the Hessian matrix.


With \hyperref[assumption_1]{Assumptions 1} and \hyperref[assumption_2]{2}, we establish the consistency of the augmented maximum likelihood estimators in the following theorem.

\begin{theorem}\label{Theorem_2.1_consistency}
Let \hyperref[assumption_1]{Assumptions 1} and \hyperref[assumption_2]{2} hold. As $N,T\rightarrow\infty$, 
\begin{align*}
    \|\widehat{\beta}-\tilde{H}\beta\|=o_P(1),
\end{align*}
where  \( \tilde{H} = \text{diag}(I_{p+1}, H^{-1}) \) and  $H=(\Lambda^\top\Lambda/N)(F^\top\tilde{F}/T)V_{NT}^{-1}$.
\end{theorem}

As established in \hyperref[Theorem_2.1_consistency]{Theorem 1}, \(\widehat{\beta}_0\) and \(\widehat{\beta}_w\) are consistent estimators of the true coefficients \(\beta_0\) and \(\beta_w\), respectively.  Moreover, \(\widehat{\beta}_f\) consistently estimates \(\beta_f\)  up to a rotation matrix \(H^{-1}\), which is a consequence of using the PCA method to recover the factors in the first stage.  This indeterminacy is well known in the factor model literature \citep[e.g.,][]{bai2003, bai2009panel}.  By Theorem 1 of \cite{baing2002}, one obtains \(\|H\| = O_P(1)\).  Under additional identification conditions on the true factors (see Conditions 2.1–2.3 in \cite{BAI201318}), the rotation matrix \(H\) converges in probability to the identity at rate  $H - I \;=\; O_P\left(\frac{1}{\sqrt{N}\wedge \sqrt{T}}\right)$.

Before deriving the asymptotic distribution of \(\widehat{\beta}\), we impose additional conditions ensuring the positive definiteness of its asymptotic covariance matrix.

\begin{assumption}\label{assumption_3}~
\begin{enumerate}
    \item $\Sigma_\beta=E\left[l_t''(\beta^\top z_t)z_tz_t^\top\right]$ is positive definite. 
    \item $\Omega_\beta=\lim_{T\rightarrow\infty}\frac{1}{T}\sum_{s,t=1}^{T-h}E\left[\xi_{t}\xi_{s}^\top\right]$ is positive definite, 
    where $\xi_{t}=\frac{(y_{t+h}-\Phi_\epsilon(\beta^{\top}z_{t}))\phi_\epsilon(\beta^{\top}z_{t})}{(1-\Phi_\epsilon(\beta^{\top}z_{t}))\Phi_\epsilon(\beta^{\top}z_{t})}z_{t}$. 
\end{enumerate}
\end{assumption}

The positive‐definiteness requirements on \(\Sigma_\beta\) and \(\Omega_\beta\) in \hyperref[assumption_3]{Assumption 3} follow directly under the i.i.d.\ case.  We now state the asymptotic distribution of \(\widehat{\beta}\) in the following theorem.

\begin{theorem}\label{Theorem_2.2_asymptotic}
Let \hyperref[assumption_1]{Assumptions 1} to \hyperref[assumption_3]{3} hold. As $N,T\rightarrow\infty$ and $\frac{\sqrt{T}}{N}\rightarrow0$, 
\begin{align*}
   \sqrt{T} (\widehat{\beta}-\tilde{H}\beta) \overset{\,\,D\,\,}{\rightarrow} \mathcal{N}(0,H_0\Sigma_\beta^{-1}\Omega_\beta\Sigma_\beta^{-1}H^\top_0),
\end{align*}
    where $H_0=\text{diag}(I_{p+1},VQ^{-1}\Sigma_\lambda^{-1})$ with $V=\text{plim}\,V_{NT}$, $Q=\text{plim}\,T^{-1}F^\top \Tilde{F}$ and $\Sigma_\lambda$ is defined in \hyperref[assumption_1]{Assumption 1}. 
\end{theorem}

In \hyperref[Theorem_2.2_asymptotic]{Theorem 2}, we derive a central limit theorem for the augmented maximum likelihood estimator \(\widehat{\beta}\).  The requirement \(\sqrt{T}/N \to 0\) controls the asymptotic order of the bias terms, ensuring that \(\widehat{\beta}\) achieves the optimal \(\sqrt{T}\)-rate of convergence.  A similar condition appears frequently in the factor model literature for panel data (see, for example, Theorem 1 of \citealp{baing2006}).

In the forecasting regression literature,  the asymptotic behaviour of fitted values is frequently examined.  For instance, Theorem 3 of \cite{baing2006} establishes the limiting distribution for predicted outcome values in linear factor‐augmented regressions.  In contrast, in binary response models, the discrete nature of \(y_{t+h}\) precludes directly forecasting the outcome itself.  Accordingly, we focus on the asymptotic behaviour of the predicted probability \(\widehat{\mathbb{P}}(y_{t+h}=1\mid \tilde{z}_{t})=\Phi_{\epsilon}(\widehat{\beta}^\top \tilde z_t)\).

\begin{theorem}\label{Theorem_2.3_asymptotic}
Let \hyperref[assumption_1]{Assumptions 1} to \hyperref[assumption_3]{3} hold. As $N,T\rightarrow\infty$ 
\begin{align*}
   \Phi_{\epsilon}(\widehat{\beta}^\top \tilde z_t)-\Phi_{\epsilon}(\beta^\top z_t) = O_P\left(\frac{1}{\sqrt{N}\wedge\sqrt{T}}\right).
\end{align*}
\end{theorem}

 \hyperref[Theorem_2.3_asymptotic]{Theorem 3} demonstrates that the proposed factor‐augmented method consistently predicts the probability that the outcome falls into one (and consequently the complementary zero category) at the rate \( O_P\left(\frac{1}{\sqrt{N}\wedge\sqrt{T}}\right)\). Under some stronger regularity conditions, one can also derive the asymptotic distribution of this predicted probability. We leave this extension for future research.

For inference, deriving analytic estimators of the asymptotic covariance matrix \(\Omega_\beta\) becomes challenging in the presence of serial dependence.  Accordingly, we follow \cite{higgins2025inference} and employ the moving‐block bootstrap (MBB) to approximate the sampling distribution of our maximum‐likelihood estimators.  Originally proposed by \cite{kunsch1989jackknife} for the dependent time series data, the MBB has recently been adapted by \cite{higgins2025inference} to accommodate panel‐data settings. This procedure randomly draws \(L\) overlapping time blocks with each of length \(q = \lfloor (T-h)/L\rfloor\) from the full sample and applies each block to all cross‐sectional units, where $\lfloor m\rfloor$ denotes the largest integer below $m$. Without loss of generality, we assume $T-h=qL$. We consider the following bootstrap procedure.

\medskip 
{\bf Step 1:} We draw $s_1, \ldots,s_L $ independently from the uniform discrete distribution on $\{0,\ldots,T-q-h\}$ and define 
\[
  y_{(l-1)q+m+h}^\ast= y_{s_l+m+h}^\ast,\, w_{(l-1)q+m}^\ast= w_{s_l+m}^\ast,\, x_{i,(l-1)q+m}^\ast= x_{i,s_l+m}^\ast,
\]
 for $l=1,\ldots,L$ and $m = 1,\ldots,q$.
 
{\bf Step 2:}  Using the bootstrap sample \(\{y^{*}_{t+h},\,w^{*}_t,\,x^{*}_{i,t}\}\), re‐estimate the model by the augmented maximum‐likelihood procedure described in Section~\ref{section-est}, yielding the bootstrap estimator \(\widehat\beta^{*}\).

{\bf Step 3:} Repeat Steps 1 and 2 a large number of times to generate the empirical bootstrap distribution of \(\widehat\beta^{*}\).

Under standard regularity conditions, the bootstrap distribution of \(\widehat\beta^{*}\) consistently approximates the sampling distribution of the original estimator \(\widehat\beta\).  We omit the formal proof, as it follows directly from existing results in the MBB literature.

\subsection{Factor number selection}\label{section-facno}

\indent
In this section, we describe our approach to determining the number of factors.  Selecting the factor dimension is a well‐studied problem in the factor‐model literature.  For instance, \cite{baing2002} propose estimating the factor count by minimizing  several information criteria, while \cite{ahn2013eigenvalue} advocate choosing the number of factors as the maximizer of the ratio of adjacent  eigenvalues.  In this paper, we follow \cite{baing2002} and employ the information criterion method to select the factor dimension.

Specifically, we consider an information criterion of the following form:
\begin{align*}
    IC(d)&=\log\left(\frac{1}{NT}\sum_{t=1}^{T}\sum_{i=1}^{N} \left[ X_{it} - \tilde{\lambda}_i^{(d)\top} \tilde{F}_t^{(d)}\right]^2\right) + d \cdot C_{NT}^{-1} \log\left(C_{NT}\right),
\end{align*}
where $\tilde{\lambda}_i^{(d)}$ and $ \tilde{F}_t^{(d)}$ represent the loading and factor estimators, respectively,  when the number of factors is specified as $d$, and $C_{NT}=\frac{NT}{N+T}$.

Let \( d_{\max} \) be a prespecified upper bound on the number of factors. The true number of factors is then estimated as
\begin{align*}
    \widehat{d}=\arg\min_{0\leq d\leq d_{\max}}\ IC(d).
\end{align*}

Using arguments similar to those in the proof of Theorem 2 and Corollary 1 of \cite{baing2002}, we can establish the consistency of this information criterion method: $\lim_{N,T\to\infty} \mathbb{P}(\widehat{d}=d_0)=1$, under \hyperref[assumption_1]{Assumption 1}.

Throughout the simulation and empirical analyses, we utilize this information criterion to select the number of factors whenever it is unknown.

\section{Simulation}\label{simulation}

In this section, we conduct simulations to examine the finite-sample performance of our proposed estimator. Specifically, we consider the following data generating process (DGP):
    \begin{align*}
    y_{t+1}=
    \begin{cases}
        1 , \  \mbox{if} \ \beta_0+\beta_f^\top f_{t}+\beta_w^\top w_{t}-\epsilon_{t+1}\geq0 \\
        0, \  \mbox{otherwise}
    \end{cases}
\end{align*}
where $\beta_0=-2$, $\beta_f=(1,1)^\top$ and $\beta_w=(1,1)^\top$. The observable factors $w_t=(w_{1t}, w_{2t})^\prime$ are generated from uniform distributions $U(0,2)$ and $U(-3,3)$, respectively, and both the unobservable factors $f_t=(f_{1t}, f_{2t})^\top$  follow AR(1) processes:
\begin{align*}
&f_{1t}=\rho_1f_{1t-1}+(1-\rho_1^2)^{1/2}\kappa_{1t},\notag\\
&f_{2t}=\rho_2f_{2t-1}+(1-\rho_2^2)^{1/2}\kappa_{2t},
\end{align*}
for $t\geq1$, with $\rho_i=0.8^i$ for $i=1,2$ and $(\kappa_{1t},\kappa_{2t})$ are independently generated from the standard normal distribution. The initial values $f_{10}$ and $f_{20}$ are independently drawn from the uniform distribution $U(0,2)$. 

Additionally, the DGP for observable predictors \{$x_{it}$\} is given by 
\[
x_{it}=\lambda_{i1} f_{1t}+\lambda_{i2} f_{2t}+\gamma_{it},
\]
where the loadings are independently generated from the uniform distribution $U(0,6)$, and $\gamma_{it}$ is following the standard normal distribution. We denote the factor estimators as $(\tilde{f}_{1t},\tilde{f}_{2t})^\top$, where $t=1,2,...,T$.

We consider two distributions for the innovations $\epsilon_{t+1}$, corresponding to light-tailed and heavy-tailed distributions, respectively. In each example, we further vary the time-series correlation of the error term, covering the i.i.d. and autocorrelation scenarios to examine the robustness of the proposed method.

\noindent
\textbf{Example 1.} Three normal distributions are considered in this example:
\begin{itemize}
\item  \textbf{DGP1:} The error term $\epsilon_t$ follows the i.i.d. standard normal distribution.
\item  \textbf{DGP2:} The error term $\epsilon_t$ is generated from the following autoregressive process:
\begin{align*}
    \epsilon_t = \rho_\epsilon \cdot \epsilon_{t-1} + \sqrt{1 - \rho_\epsilon^2} \cdot \nu_t,
\end{align*}
where $\rho_\epsilon = 0.3$ and $\nu_t$ follows the i.i.d. standard normal distribution.

\item  \textbf{DGP3:} $ \epsilon_t$ follows the same  autoregressive process as DGP2 with $\rho_\epsilon = 0.7$.
\end{itemize}

\noindent
\textbf{Example 2.} We adopt DGPs analogous to those in {\bf Example 1}, replacing every normal distribution with the logistic distribution with zero mean and unit variance.

For each simulation dataset, we employ the method outlined in Section 2 to construct the estimators and predicted probabilities. We vary the values of $N$ and $T$ from $\{100,200,300\}$ and $\{100, 200,400\}$, respectively.  After repeating the procedure for $R=500$ times, we compute the root mean squared errors (RMSEs) to evaluate the finite-sample estimation accuracy of the proposed method:
\[
{\rm RMSE}_{all}=\sqrt{\frac{1}{R}\sum_{i=1}^{R}\| \widehat{\beta}_i-\beta^\circ\|^2},\,
\]
where $\widehat{\beta}_i$ denotes the estimated coefficient vector in the $i$th replication, and $\beta^\circ$ is the rotated coefficeints as defined in Section \ref{section-asy}. Additionally, We also report RMSE values for each individual coefficient, including RMSE$_{\text{cons}}$, RMSE$_{f_1}$, RMSE$_{f_2}$, RMSE$_{w_1}$, and RMSE$_{w_2}$.


We compute the area under the receiver operating characteristic curve (AUC) to assess predictive performance. The AUC measures the model's ability to distinguish between the two outcome classes (0 vs 1) across different threshold values. The receiver operating characteristic curve (ROC) curve maps the true positive rate
\begin{equation*}
\operatorname{TP}(\xi)=P_{t-h}\left(p_t>\xi \mid y_t=1\right)
\end{equation*}
against the false positive rate
\begin{equation*}
\operatorname{FP}(\xi)=P_{t-h}\left(p_t>\xi \mid y_t=0\right)
\end{equation*}
across all thresholds $0 \leq \xi \leq 1$, forming a function over the unit square $[0,1] \times [0,1]$. Here, $p_t$ denotes the predicted conditional probability of the target variable. And a higher AUC indicates stronger predictive discrimination. Specifically, for each replication, we calculate the AUC based on the estimated probabilities and the true binary outcomes. The simulation results are then summarized by reporting the mean, median, and standard deviation of the AUC across all replications.

Table \ref{tab:simulation1} and Table \ref{tab:simulation2} present the simulation results for the two Examples. The findings indicate that our method delivers robust performance under both light-tailed and heavy-tailed error distributions. The RMSE values decrease steadily as the sample size \(T\) increases. This pattern supports the \(\sqrt{T}\)-consistency of the augmented maximum likelihood estimators established in Theorem \ref{Theorem_2.2_asymptotic}.

Tables \ref{tab:simulation3} and \ref{tab:simulation4} summarize the mean, median, and standard deviation of the resulting AUCs. Overall, the proposed model exhibits high predictive accuracy, with both the mean and median AUC values exceeding 0.89 across simulations under various settings.  Comparisons between Tables~\ref{tab:simulation3} and~\ref{tab:simulation4} reveal that variations in the distribution of idiosyncratic error terms have limited influence on the AUC results.


\begin{landscape}
\begin{table}[H]
\centering
\caption{RMSEs for  Example 1.}
\label{tab:simulation1}
\resizebox{1.17\textwidth}{!}{
\begin{tabular}{lllllllllllllllllll}
\hline
    & \multicolumn{3}{l}{RMSE$_{\text{all}}$} & \multicolumn{3}{l}{RMSE$_{\text{cons}}$} & \multicolumn{3}{l}{RMSE$_{f_1}$} & \multicolumn{3}{l}{RMSE$_{f_2}$} & \multicolumn{3}{l}{RMSE$_{w_1}$} & \multicolumn{3}{l}{RMSE$_{w_2}$} \\ \hline
N/T & 100    & 200    & 400    & 100      & 200     & 400     & 100      & 200     & 400     & 100      & 200     & 400     & 100      & 200     & 400     & 100      & 200     & 400     \\ \hline
    & \multicolumn{18}{l}{\emph{Panel A: DGP1}}                                                                                                                                                           \\
100 & 1.063  & 0.632  & 0.423  & 0.772    & 0.463   & 0.310   & 0.460    & 0.266   & 0.175   & 0.263    & 0.151   & 0.112   & 0.399    & 0.247   & 0.163   & 0.307    & 0.173   & 0.114   \\
200 & 1.051  & 0.625  & 0.406  & 0.755    & 0.448   & 0.289   & 0.474    & 0.273   & 0.180   & 0.239    & 0.158   & 0.110   & 0.399    & 0.239   & 0.155   & 0.305    & 0.184   & 0.112   \\
300 & 1.054  & 0.620  & 0.398  & 0.763    & 0.449   & 0.288   & 0.460    & 0.272   & 0.164   & 0.248    & 0.157   & 0.106   & 0.399    & 0.234   & 0.157   & 0.311    & 0.172   & 0.113   \\ \hline
    & \multicolumn{18}{l}{\emph{Panel B: DGP2}}                                                                                                                                                           \\
100 & 1.038  & 0.704  & 0.424  & 0.727    & 0.512   & 0.304   & 0.472    & 0.309   & 0.188   & 0.286    & 0.172   & 0.115   & 0.383    & 0.262   & 0.154   & 0.314    & 0.196   & 0.121   \\
200 & 1.088  & 0.645  & 0.428  & 0.784    & 0.465   & 0.313   & 0.493    & 0.271   & 0.177   & 0.252    & 0.169   & 0.118   & 0.404    & 0.251   & 0.166   & 0.314    & 0.188   & 0.113   \\
300 & 1.115  & 0.649  & 0.437  & 0.810    & 0.461   & 0.322   & 0.497    & 0.294   & 0.182   & 0.266    & 0.171   & 0.113   & 0.414    & 0.244   & 0.166   & 0.315    & 0.183   & 0.120   \\ \hline
    & \multicolumn{18}{l}{\emph{Panel C: DGP3}}                                                                                                                                                           \\
100 & 1.227  & 0.744  & 0.474  & 0.887    & 0.549   & 0.341   & 0.550    & 0.319   & 0.215   & 0.316    & 0.197   & 0.131   & 0.442    & 0.266   & 0.169   & 0.353    & 0.202   & 0.126   \\
200 & 1.169  & 0.757  & 0.461  & 0.842    & 0.542   & 0.338   & 0.524    & 0.346   & 0.192   & 0.304    & 0.203   & 0.133   & 0.411    & 0.271   & 0.169   & 0.350    & 0.213   & 0.124   \\
300 & 1.259  & 0.740  & 0.473  & 0.918    & 0.532   & 0.342   & 0.567    & 0.335   & 0.205   & 0.319    & 0.202   & 0.132   & 0.437    & 0.264   & 0.170   & 0.359    & 0.205   & 0.134   \\ \hline
\end{tabular}
}
\end{table}
\vspace{1cm}
\begin{table}[H]
\centering
\caption{RMSEs for Example 2.}
\label{tab:simulation2}
\resizebox{1.17\textwidth}{!}{
\begin{tabular}{lllllllllllllllllll}
\hline
    & \multicolumn{3}{l}{RMSE$_{\text{all}}$} & \multicolumn{3}{l}{RMSE$_{\text{cons}}$} & \multicolumn{3}{l}{RMSE$_{f_1}$} & \multicolumn{3}{l}{RMSE$_{f_2}$} & \multicolumn{3}{l}{RMSE$_{w_1}$} & \multicolumn{3}{l}{RMSE$_{w_2}$} \\ \hline
N/T & 100    & 200    & 400    & 100      & 200     & 400     & 100      & 200     & 400     & 100      & 200     & 400     & 100      & 200     & 400     & 100      & 200     & 400     \\ \hline
    & \multicolumn{18}{l}{\emph{Panel A: DGP1}}                                                                                                                                                           \\
100 & 1.178  & 0.693  & 0.482  & 0.858    & 0.499   & 0.352   & 0.474    & 0.285   & 0.189   & 0.331    & 0.220   & 0.151   & 0.471    & 0.264   & 0.192   & 0.309    & 0.177   & 0.114   \\
200 & 1.163  & 0.720  & 0.489  & 0.825    & 0.521   & 0.349   & 0.504    & 0.294   & 0.202   & 0.362    & 0.214   & 0.151   & 0.439    & 0.283   & 0.196   & 0.311    & 0.187   & 0.122   \\
300 & 1.155  & 0.688  & 0.462  & 0.836    & 0.499   & 0.331   & 0.497    & 0.283   & 0.191   & 0.325    & 0.214   & 0.147   & 0.444    & 0.258   & 0.185   & 0.292    & 0.178   & 0.108   \\ \hline
    & \multicolumn{18}{l}{\emph{Panel B: DGP2}}                                                                                                                                                           \\
100 & 1.196  & 0.717  & 0.475  & 0.824    & 0.512   & 0.338   & 0.563    & 0.300   & 0.198   & 0.366    & 0.250   & 0.165   & 0.448    & 0.260   & 0.179   & 0.315    & 0.176   & 0.113   \\
200 & 1.272  & 0.714  & 0.485  & 0.906    & 0.517   & 0.347   & 0.599    & 0.287   & 0.201   & 0.339    & 0.230   & 0.166   & 0.460    & 0.278   & 0.186   & 0.334    & 0.172   & 0.109   \\
300 & 1.167  & 0.747  & 0.520  & 0.823    & 0.532   & 0.381   & 0.517    & 0.313   & 0.202   & 0.380    & 0.254   & 0.168   & 0.439    & 0.273   & 0.201   & 0.284    & 0.195   & 0.124   \\ \hline
    & \multicolumn{18}{l}{\emph{Panel C: DGP3}}                                                                                                                                                           \\
100 & 1.376  & 0.858  & 0.534  & 0.982    & 0.622   & 0.379   & 0.641    & 0.366   & 0.236   & 0.436    & 0.311   & 0.194   & 0.466    & 0.284   & 0.179   & 0.333    & 0.193   & 0.124   \\
200 & 1.452  & 0.830  & 0.560  & 1.033    & 0.605   & 0.408   & 0.690    & 0.364   & 0.238   & 0.443    & 0.271   & 0.194   & 0.498    & 0.285   & 0.196   & 0.346    & 0.191   & 0.120   \\
300 & 1.342  & 0.910  & 0.577  & 0.942    & 0.676   & 0.425   & 0.624    & 0.388   & 0.242   & 0.438    & 0.300   & 0.200   & 0.465    & 0.294   & 0.194   & 0.339    & 0.209   & 0.126   \\ \hline
\end{tabular}
}
\end{table}
\end{landscape}

\noindent
\begin{table}[H]
\centering
\caption{AUCs for Example 1.}
\label{tab:simulation3}
\vspace{0.3em}
\rotatebox{0}{
\resizebox{0.5\textheight}{!}{
\begin{tabular}{llllllllll}
\hline
    & \multicolumn{3}{l}{AUC$_{\text{mean}}$} & \multicolumn{3}{l}{AUC$_{\text{median}}$} & \multicolumn{3}{l}{AUC$_{\text{std}}$} \\ \hline
N/T & 100    & 200    & 400    & 100      & 200     & 400     & 100      & 200     & 400     \\ \hline
    & \multicolumn{9}{l}{\emph{Panel A: DGP1}} \\
100 & 0.963  & 0.960  & 0.958  & 0.965    & 0.961   & 0.958   & 0.017    & 0.011   & 0.009   \\
200 & 0.963  & 0.959  & 0.959  & 0.965    & 0.960   & 0.959   & 0.017    & 0.013   & 0.008   \\
300 & 0.961  & 0.960  & 0.958  & 0.963    & 0.961   & 0.959   & 0.018    & 0.012   & 0.008   \\ \hline
    & \multicolumn{9}{l}{\emph{Panel B: DGP2}} \\
100 & 0.962  & 0.961  & 0.958  & 0.965    & 0.962   & 0.959   & 0.016    & 0.012   & 0.009   \\
200 & 0.963  & 0.961  & 0.959  & 0.965    & 0.961   & 0.959   & 0.016    & 0.012   & 0.009   \\
300 & 0.963  & 0.960  & 0.958  & 0.964    & 0.961   & 0.958   & 0.018    & 0.012   & 0.009   \\ \hline
    & \multicolumn{9}{l}{\emph{Panel C: DGP3}} \\
100 & 0.904  & 0.900  & 0.894  & 0.907    & 0.901   & 0.894   & 0.034    & 0.023   & 0.018   \\
200 & 0.906  & 0.898  & 0.896  & 0.910    & 0.899   & 0.897   & 0.035    & 0.026   & 0.017   \\
300 & 0.907  & 0.901  & 0.895  & 0.911    & 0.901   & 0.895   & 0.034    & 0.023   & 0.018   \\ \hline
\end{tabular}
}
}
\end{table}

\vspace{1em} 

\begin{table}[H]
\centering
\caption{AUCs for Example 2.}
\label{tab:simulation4}
\vspace{0.3em}
\rotatebox{0}{
\resizebox{0.5\textheight}{!}{
\begin{tabular}{llllllllll}
\hline
    & \multicolumn{3}{l}{AUC$_{\text{mean}}$} & \multicolumn{3}{l}{AUC$_{\text{median}}$} & \multicolumn{3}{l}{AUC$_{\text{std}}$} \\ \hline
N/T & 100    & 200    & 400    & 100      & 200     & 400     & 100      & 200     & 400     \\ \hline
    & \multicolumn{9}{l}{\emph{Panel A: DGP1}} \\
100 & 0.963  & 0.960  & 0.958  & 0.965    & 0.961   & 0.958   & 0.017    & 0.011   & 0.009   \\
200 & 0.963  & 0.959  & 0.959  & 0.965    & 0.960   & 0.959   & 0.017    & 0.013   & 0.008   \\
300 & 0.961  & 0.960  & 0.958  & 0.963    & 0.961   & 0.959   & 0.018    & 0.012   & 0.008   \\ \hline
    & \multicolumn{9}{l}{\emph{Panel B: DGP2}} \\
100 & 0.962  & 0.961  & 0.958  & 0.965    & 0.962   & 0.959   & 0.016    & 0.012   & 0.009   \\
200 & 0.963  & 0.961  & 0.959  & 0.965    & 0.961   & 0.959   & 0.016    & 0.012   & 0.009   \\
300 & 0.963  & 0.960  & 0.958  & 0.964    & 0.961   & 0.958   & 0.018    & 0.012   & 0.009   \\ \hline
    & \multicolumn{9}{l}{\emph{Panel C: DGP3}} \\
100 & 0.904  & 0.900  & 0.894  & 0.907    & 0.901   & 0.894   & 0.034    & 0.023   & 0.018   \\
200 & 0.906  & 0.898  & 0.896  & 0.910    & 0.899   & 0.897   & 0.035    & 0.026   & 0.017   \\
300 & 0.907  & 0.901  & 0.895  & 0.911    & 0.901   & 0.895   & 0.034    & 0.024   & 0.018   \\ \hline
\end{tabular}
}
}
\end{table}

\newpage
\section{Empirical Study}\label{a_case_study}

Recession forecasting is a critical task undertaken by many economic institutions. Predicting whether the economy will enter an expansion or recession in the coming months or years provides valuable insights for policymakers, investors, and households. For instance, government agencies can adjust fiscal policies based on projected recovery timelines, while central banks can refine monetary strategies in anticipation of future business cycle shifts. Seminal studies, such as \citet{estrella1998predicting,chauvet2005forecasting} and \citet{kauppi2008predicting}, demonstrate the predictive power of yield curve inversions and interest rate spread for U.S. recessions, using dynamic Probit models. An expanding body of research continues to identify additional predictive variables, including sentiment indicators \citep{christiansen2014forecasting}, economic uncertainty measures \citep{ercolani2020forecasting}, and credit market variables \citep{ng2012forecasting,ponka2017role}. Meanwhile, researchers have increasingly focused on employing high-dimensional information and more advanced methods to predict U.S. recessions \citep[see,][]{fornaro2016forecasting,berge2015predicting,liu2016predicts,vrontos2021modeling,proano2014predicting}. Building on this literature, we conduct an empirical study to evaluate the proposed binary FAR model in predicting U.S. recession periods.

\subsection{U.S. recessions}
Following the same definition of recessions as \citet{estrella2006yield}, we use the NBER defined business cycle expansion and contraction dates to determine US recessions. In particular, the first month following a peak month defines the first recession month and the last month of a through defines the last recession month. Let $y_t$ denote the state of the business cycle as determined by the NBER, where $y_t=1$ denotes that month $t$ is an NBER-defined recession and $y_t=0$ is an expansion. That is,
\begin{equation*}
y_t= \begin{cases}1, & \text { if the economy is in a recession at time } $t$, \\ 0, & \text { if the economy is in an expansion at time } $t$.\end{cases}
\end{equation*}

A crucial limitation arises from the NBER's ex post dating procedure, which introduces substantial reporting delays. We follow \citet{chauvet2005forecasting,kauppi2008predicting,ng2012forecasting} and \citet{christiansen2014forecasting} to assume that the recession indicator is announced with a three month delay, as they argue that the recent advances in real-time business cycle dating let researchers make reasonable assumptions on the state of the economy even if the NBER is yet to announce its classification. Specifically, we estimate model parameters using data measured at time $t - d - q$ (rather than $t - q$) for prediction construction, where $d=3$ denotes the assumed publication lag of the recession indicator to account for the announcement delay.

\subsection{Data}
The NBER business cycle dates are publicly available and are standard in the business cycle literature.\footnote{The business cycle dates are available at \url{www.nber.org.}} We use a large, monthly frequency, macroeconomic dataset from which we estimate the monthly
macroeconomic factors: 121 variables included in the FRED-MD database described in \citet{mccracken2016fred}. The series represent broad categories of macroeconomic and financial time series including: output and income; labor market; housing; consumption, orders, and inventories; money and credit; interest and exchange rates; prices; and stock market. For detailed descriptions of these macroeconomic datasets, please refer to Michael W. McCracken's website\footnote{\url{https://research.stlouisfed.org/econ/mccracken/fred-databases/}}. We follow \citet{mccracken2016fred}, using the transform precodure they offer to ensure the stationarity of the dataset. Our sample period spans from January 1960 to December 2024, providing a total of $T=780$ monthly observations. The sample period includes nine recession periods, with the most recent occurring during the COVID-19 pandemic from March to April 2020.

\subsection{Common factors} \label{sec:common_factors}
The $IC_2$ criterion \citep{baing2002} indicates that the optimal number of factors is 8. To better understand how these unobservable common factors relate to observable variables, we identify which series they best explain. Following \citet{mccracken2016fred}, we regress each $i$-th series in the transformed dataset on a set of $k$ stationary common factors. This yields the adjusted $R_i^2(r)$ for each series $i$, where $r$ ranges from $1$ to $k$. Since the factors are orthogonal and ordered by decreasing eigenvalues, we can determine the incremental explanatory power of factor $r$ for series $i$ using the ``marginal $R$-squared" ($mR^2$). This is calculated as $mR^2_i(r) = R^2_i(r) - R^2_i(r-1)$ for $r \ge 2$, with $mR^2_i(1) = R_i^2(1)$. The average importance of factor $r$ across all series is then $mR^2(r) = \frac{1}{N}\sum_{i=1}^N mR_i^2(r)$.

Table \ref{tab:mR2} presents the $mR^2(r)$ values and lists the five series with the highest $mR^2_i(r)$ for each factor $r$. Typically, the observable series with the highest $mR^2_i(r)$ for a given factor are similar and belong to the same category. This allows us to assign economic meaning to each common factor based on its corresponding observable series. Furthermore, we select the observable series with the highest $mR^2(r)$ within its category to serve as an observable representative for each macroeconomic factor, which we refer to as the ``observable factor" (bolded in Table \ref{tab:mR2}). These eight factors collectively explain 52.3\% of the total variation across all series. They are categorized as: two output factors, a prices factor, three interest factors, a labor factor, and a stock market factor. The corresponding observable factors are: IP: Manufacturing (SIC) index (\textbf{IPMANSICS}), CPI index: Commodities (\textbf{CPIAUCSL}), Baa-FF spread (\textbf{BAAFFM}), 1-year treasury rate (\textbf{GS1}), 5-year treasury spread (\textbf{T5YFFM}), average weekly working hours: manufacturing (\textbf{AWHMAN}), Real Personal Income (\textbf{RPI}), and S\&P 500 index (\textbf{S\&P 500}).

\begin{table}[H]
    \onehalfspacing
    \centering
    \scriptsize
    \caption{Marginal explain power of common factors: Total variation explained, 52.3\%.}
    \label{tab:mR2}
    \begin{threeparttable}
    \begin{tabular}{llllllll}
    \hline
                  & $mR^2$  &                  & $mR^2$   &                 & $mR^2$ &                  & $mR^2$  \\ \hline
    \emph{Factor 1: Output}      & 0.192                          & \emph{Factor 2: Prices}         & 0.076                           & \emph{Factor 3: Interest}        & 0.068                          & \emph{Factor 4: Interest}         & 0.055                         \\
                  \textbf{IPMANSICS}     & 0.827                          & \textbf{CUSR0000SAC}      & 0.531                           & \textbf{BAAFFM}          & 0.347                          & \textbf{GS1}              & 0.585                         \\
    PAYEMS        & 0.814                          & CUSR0000SA0L2    & 0.518                           & AAAFFM          & 0.336                          & TB6MS              & 0.540                         \\
    USGOOD        & 0.807                          & CPIAUCSL  & 0.492                           & HOUST         & 0.327                          & GS5            & 0.523                         \\
    IPFPNSS        & 0.791                          & DNDGRG3M086SBEA         & 0.491                           & T10YFFM          & 0.300                          & TB3MS             & 0.462                         \\
    INDPRO       & 0.784                          & CPITRNSL    & 0.485                           & PERMIT        & 0.287                          & GS10            & 0.456                         \\ \hline
    \emph{Factor 5: Interest}      & 0.050                          & \emph{Factor 6: Labor}         & 0.032                           & \emph{Factor 7: Output}        & 0.025                          & \emph{Factor 8: Stock}         & 0.025                \\
    \textbf{T5YFFM}      & 0.369                          & \textbf{AWHMAN}           & 0.254                           & \textbf{RPI}    & 0.247                          & \textbf{S\&P 500}           & 0.238                         \\
    T1YFFM         & 0.345                          & CES0600000007    & 0.232                           & CONSPI        & 0.229                          & S\&P div yield        & 0.158                         \\
    TB6SMFFM        & 0.321                          & UEMP15OV            & 0.225                           & CES0600000007  & 0.114                          & UEMP15OV         & 0.134                         \\
    PERMIT        & 0.319                          & S\&P PE ratio          & 0.217                           & AWHMAN        & 0.108                          & S\&P PE ratio              & 0.114                         \\
    T10YFFM      & 0.308                          & S\&P 500           & 0.127                           & IPMAT       & 0.093                          & CES0600000007          & 0.096                         \\              \hline
    \end{tabular}
    \begin{tablenotes}[para,flushleft]
              \emph{Note}: This table presents the marginal explain power of common factors. The factors are assigned economic meaning according to the five series with the highest $mR^2_i(r)$ listed in the table. The number next to each factor is the average marginal $R$-squared. The bolded series are the observable representatives for each common factor, namely, the observable factor.
            \end{tablenotes}
        \end{threeparttable}
    \end{table}

\subsection{Models}

To have a clear view of the performance of the proposed binary FAR model, we consider two models in our empirical study.
\begin{itemize}
    \item Binary FAR. \begin{align*}
    &y_{t+h}=
    \begin{cases}
        1 , \  \mbox{if} \ \beta^{\top}z^f_{t}-\epsilon_{t+h}\geq0, \\
        0, \  \mbox{otherwise},
    \end{cases}\\
    &X_t=\Lambda F_t+e_t,
\end{align*}
where $y_t$ is the U.S. recession dummy variable, $f_t$ is the common factors, $z_t^f = (1,f_t^\top)^\top$, and $X=(X_1,\ldots,X_N)^\top$ denotes the high-dimensional observable dataset.
\item Probit regression. \begin{equation*}
    y_{t+h}=
    \begin{cases}
        1 , \  \mbox{if} \ \beta^{\top}z^w_{t}-\epsilon_{t+h}\geq0, \\
        0, \  \mbox{otherwise},
    \end{cases}
\end{equation*}
where $y_t$ is the U.S. recession dummy variable and $w_t$ is a vector of observable variables, $z_t^w = (1,w_t^\top)^\top$.
\end{itemize}

In our empirical study, $w_t$ contains the eight observable factors in section \ref{sec:common_factors}, that is, $w_t=(IPMANSICS_t,CPIAUCSL_t,BAAFFM_t,GS1_t,T5YFFM_t,AWHMAN_t,RPI_t,$ $S\&P 500_t)^\top$. All series are transformed to be stationary, for example, $S\&P 500_t$ denotes the log return of S\&P 500 index. Given the limitations of traditional Probit models in handling high-dimensional settings, we assess the merits of the binary FAR model-based on latent factors—by contrasting its performance with that of a Probit regression relying on observable proxies for the same factors. Our analysis considers $h=1, 3, 6, 9, 12$, corresponding to 1-month, 1-quarter, 2-quarter, 3-quarter and 1-year-ahead predictions, respectively.

\subsection{In-sample results}
In this section, we use the full sample from 1960:M1 to 2024:M12 to evaluate the in-sample fitting performance of binary FAR and traditional Probit regression. Two goodness-of-fit measures are considered, the first is the area under the receiver operating characteristic curve (AUC), as introduced in section \ref{simulation}. The second measure is the pseudo-$R^2$ of \citet{estrella1998new}, which is a counterpart of the coefficient of determination ($R^2$) designed for binary response models, given by
\begin{equation*}
\operatorname{Pseudo}-R^2=1-\left(\frac{\log L_u}{\log L_c}\right)^{-(2 / T) \log L_c}
\end{equation*}
where $\log L_u$ and $\log L_c$ are the maximum values of the constrained and unconstrained log-likelihood functions respectively, and $T$ is the sample size. This measure takes on values between 0 and 1, and can be interpreted in the same way as the coefficient of determination in the usual linear predictive regression model.

\begin{table}[H]
\centering
\caption{In-sample results.}
\label{tab:is-results}
\begin{tabular}{llllll}
\hline
                  & $h=1$   & $h=3$   & $h=6$   & $h=9$   & $h=12$  \\ \hline
                  & \multicolumn{5}{l}{AUC}               \\ \cline{2-6}
Binary FAR        & 0.962 & 0.979 & 0.942 & 0.933 & 0.919 \\
Probit regression & 0.948 & 0.963 & 0.917 & 0.915 & 0.911 \\ \hline
                  & \multicolumn{5}{l}{Pseudo-$R^2$}         \\ \cline{2-6}
Binary FAR        & 0.462 & 0.494 & 0.381 & 0.274 & 0.244 \\
Probit regression & 0.432 & 0.462 & 0.305 & 0.266 & 0.251 \\ \hline
\end{tabular}
\end{table}

\begin{figure}[H]
    \centering
    \begin{subcaptionbox}{Binary FAR\label{fig:is-roc-bfar}}[0.45\linewidth]
        {\includegraphics[width=\linewidth]{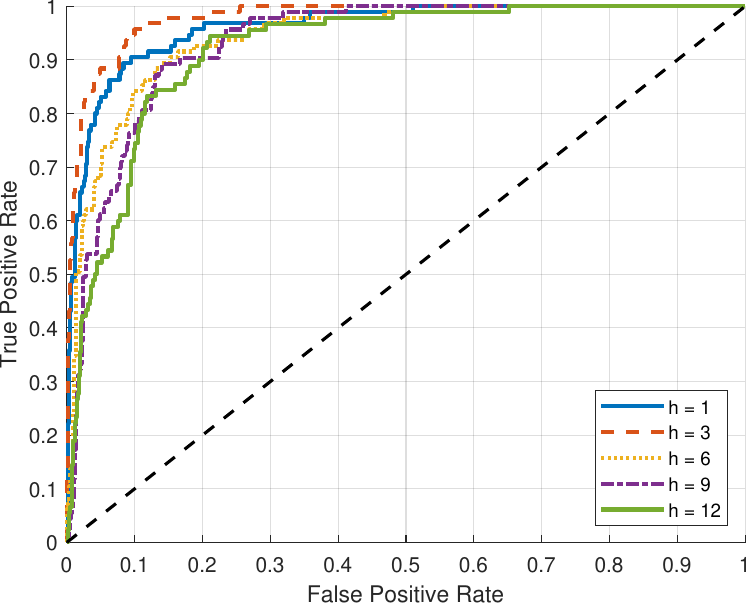}}
    \end{subcaptionbox}
    \hfill
    \begin{subcaptionbox}{Probit regression\label{fig:is-roc-prob}}[0.45\linewidth]
        {\includegraphics[width=\linewidth]{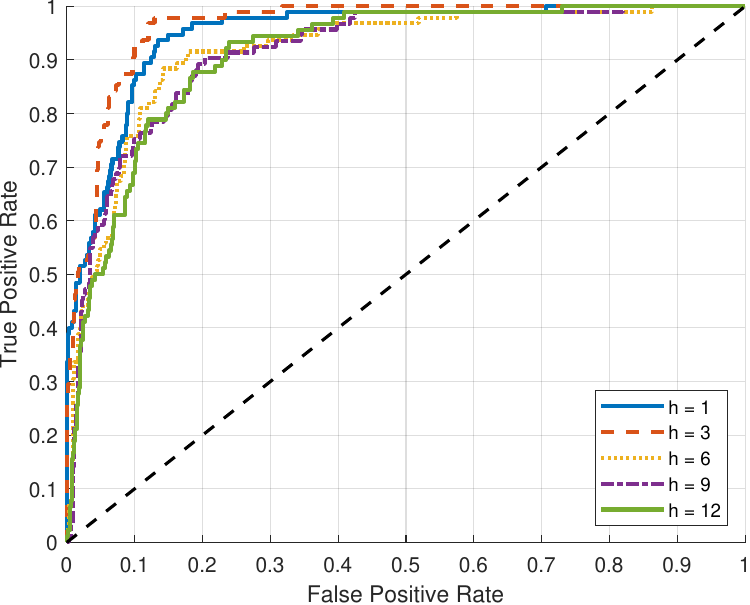}}
    \end{subcaptionbox}
    \caption{In-sample ROC curves. This figure plots in-sample ROC curves for binary FAR and Probit regression. The 45° line represents a coin-toss classifier.}
    \label{fig:is-roc}
\end{figure}
       
Table \ref{tab:is-results} presents the in-sample performance of the binary FAR and Probit regression models across multiple forecast horizons ($h = 1, 3, 6, 9, 12$). Both models shows strong discriminatory power, as measured by the AUC, and moderate explanatory power, as indicated by the Pseudo-$R^2$. Across all horizons, binary FAR consistently outperforms Probit regression, with the performance gap being especially pronounced at shorter horizons. Figure \ref{fig:is-roc} plots the corresponding in-sample ROC curves, which confirm  the findings in Table \ref{tab:is-results}.  A curve lying above the 45$^\circ$ diagonal indicates classification performance better than random guessing. Notably, the binary FAR model demonstrates superior classification performance, while the Probit regression model, which relies solely on observable factors, shows weaker discriminatory capacity.

\begin{figure}[H]
    \centering
    \begin{subfigure}[b]{\textwidth}
    \centering
    \includegraphics[width=0.18\linewidth]{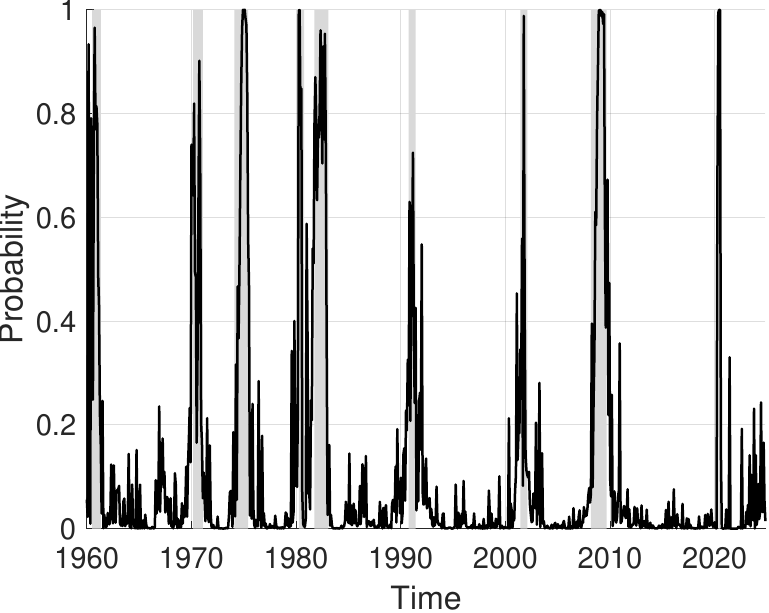}
    \includegraphics[width=0.18\linewidth]{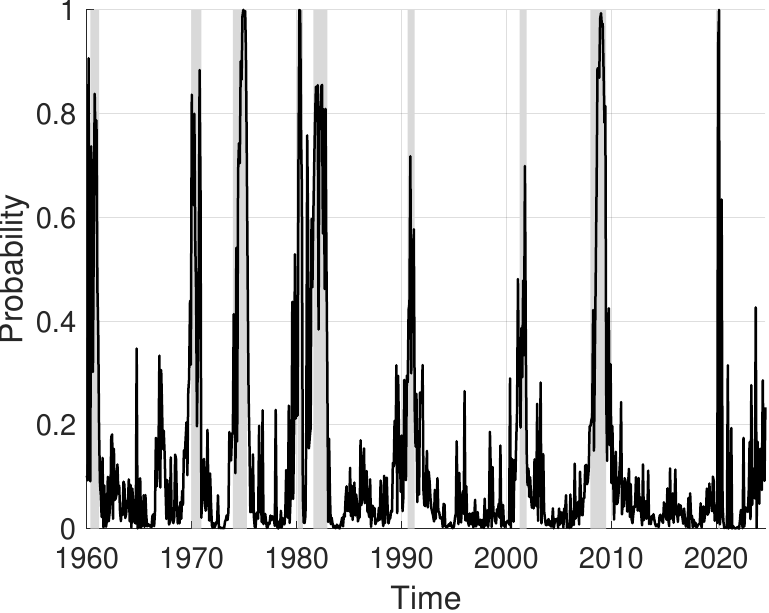}
    \includegraphics[width=0.18\linewidth]{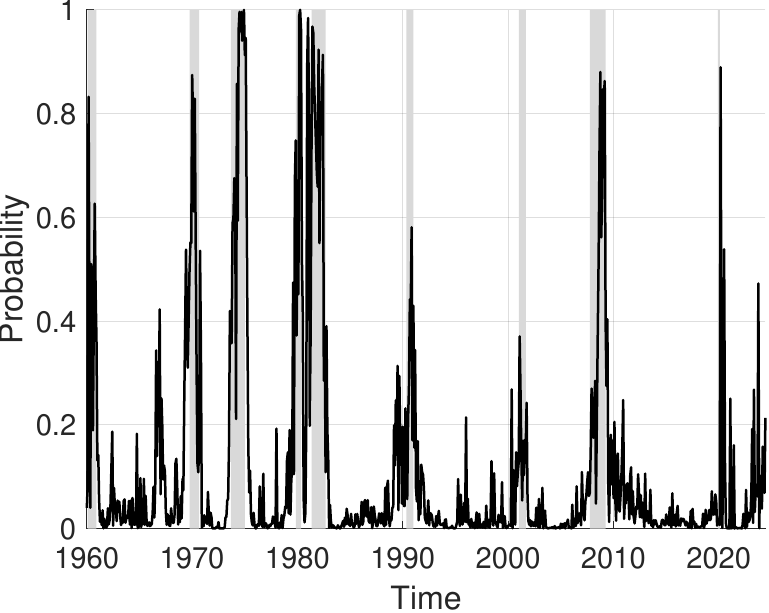}
    \includegraphics[width=0.18\linewidth]{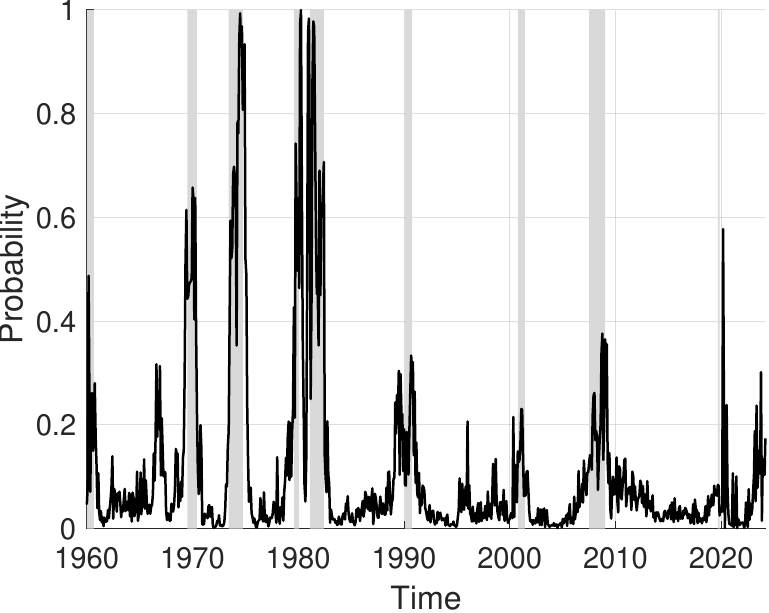}
    \includegraphics[width=0.18\linewidth]{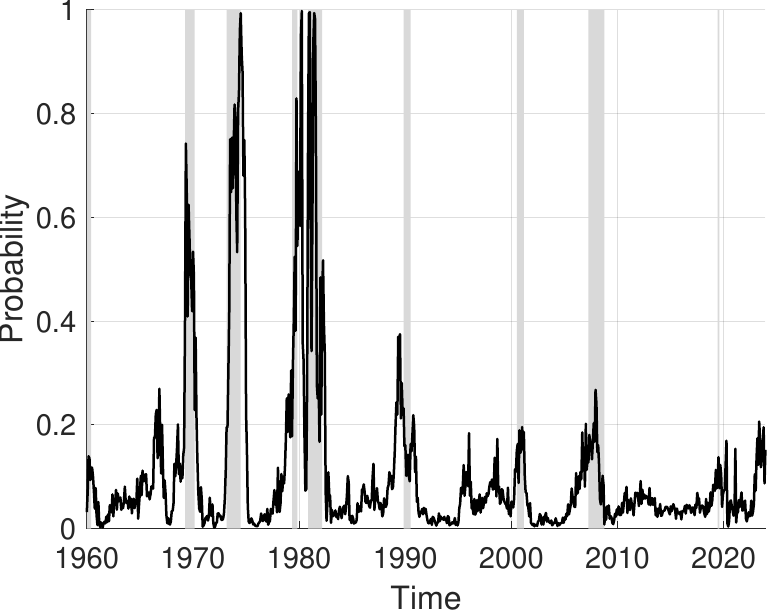}
    \caption{Binary FAR.} 
    \end{subfigure}
    
     \begin{subfigure}[b]{\textwidth}
        \centering
        \begin{minipage}[b]{0.18\linewidth}
            \centering
            \includegraphics[width=\linewidth]{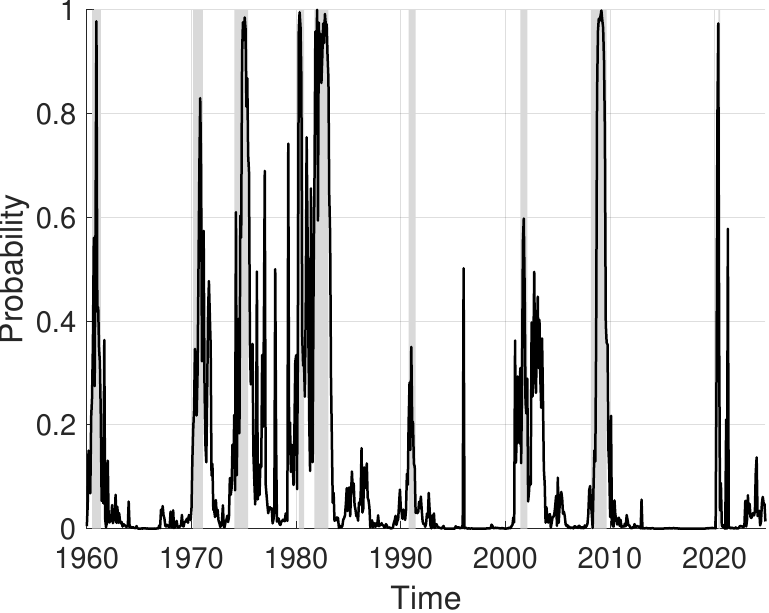}
            \small $h=1$
        \end{minipage}
        \begin{minipage}[b]{0.18\linewidth}
            \centering
            \includegraphics[width=\linewidth]{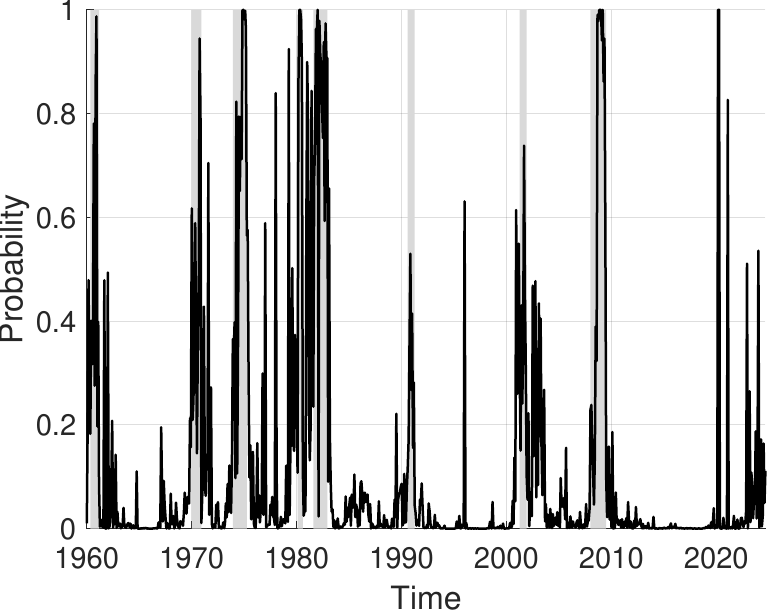}
            \small $h=3$
        \end{minipage}
        \begin{minipage}[b]{0.18\linewidth}
            \centering
            \includegraphics[width=\linewidth]{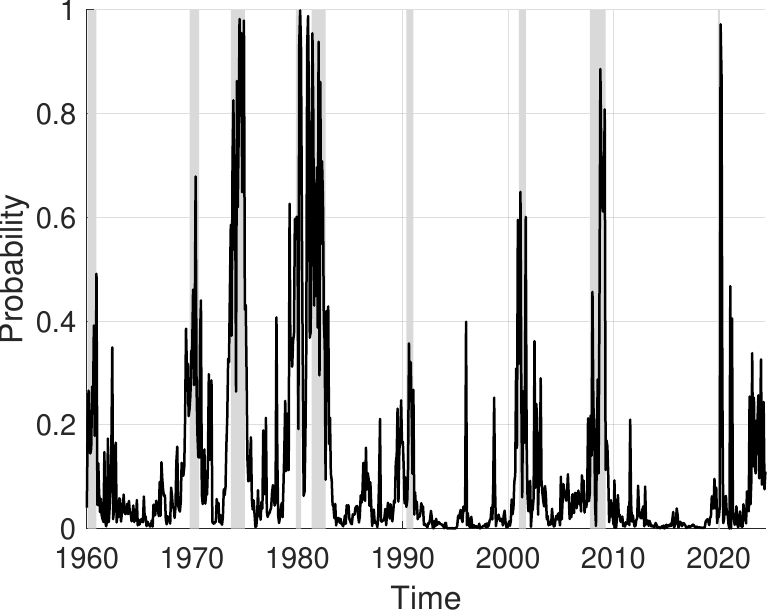}
            \small $h=6$
        \end{minipage}
        \begin{minipage}[b]{0.18\linewidth}
            \centering
            \includegraphics[width=\linewidth]{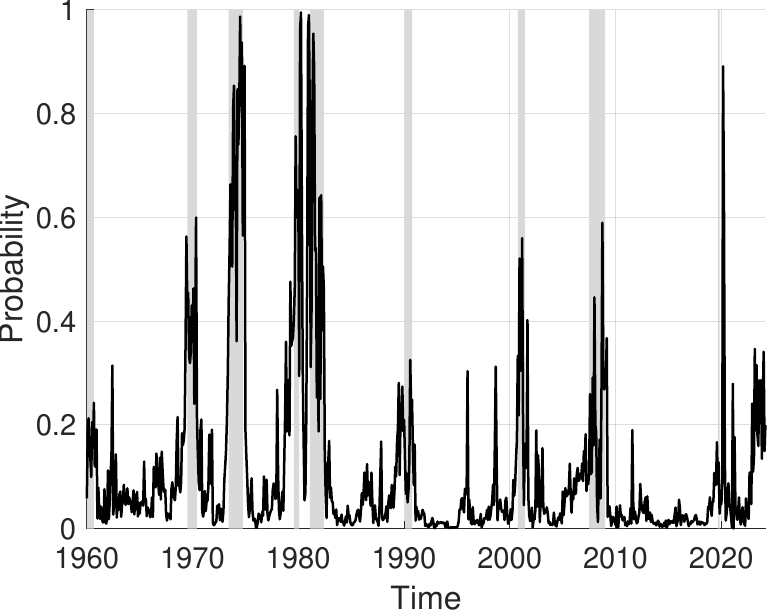}
            \small $h=9$
        \end{minipage}
        \begin{minipage}[b]{0.18\linewidth}
            \centering
            \includegraphics[width=\linewidth]{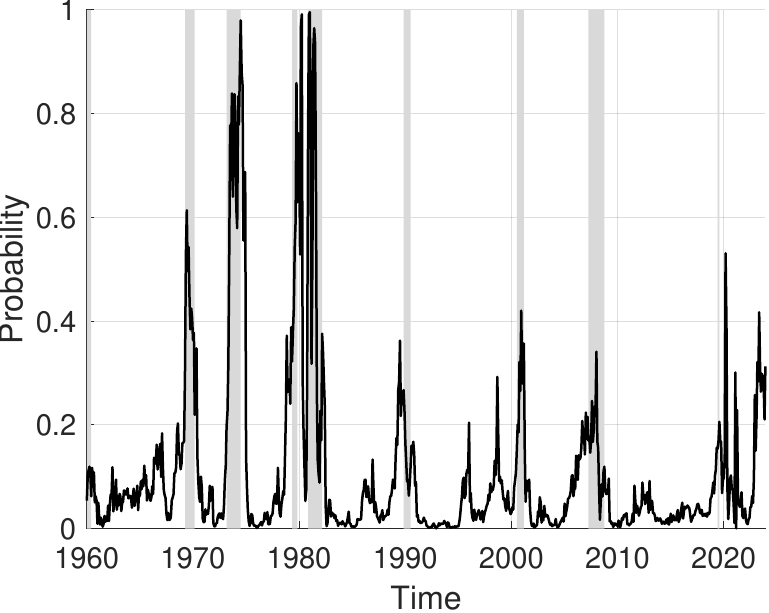}
            \small $h=12$
        \end{minipage}
        \caption{Probit regression.} 
    \end{subfigure}

    \caption{In-sample fit. This figure plots in-sample implied recession probabilities. NBER defined recession dates are in grey shading.}
    \label{fig:is-fit}
\end{figure}

Figure \ref{fig:is-fit} contains the in-sample implied recession probability for binary FAR and Probit regression with NBER defined recession dates plotted in grey shading. The figure demonstrates that binary FAR consistently generates higher fitted probabilities during recession periods compared to Probit regression, particularly at shorter horizons. This superior performance is especially evident during the early 1990s and early 2000s recessions.

To sum up, the in-sample results demonstrate that the binary FAR model, by leveraging high-dimensional information in latent factors, provides significantly richer information about future business cycle periods than the conventional low-dimensional Probit regression.

\subsection{Out-of-sample results}

It is well-known that in-sample results may not reliably reflect out-of-sample predictive accuracy. In order to evaluate the predictive accuracy of our models, we conduct an out-of-sample analysis. In particular, the full sample is split into an initial estimation period from 1960:M1 to 1999:M12 and an out-of-sample forecasting period from 2000:M1 to 2024:M12, yielding a total of $T = 300$ out-of-sample forecasts. The out-of-sample period includes the three major U.S. recessions: the dot-com crisis in 2001, the subprime crisis in 2008 and the recent COVID-19 pandemic in 2020. We employ an expanding window approach, whereby the estimation sample grows by one observation with each step forward in time. Since the Pseudo-$R^2$ is an in-sample measure, we use the out-of-sample AUC as the sole criterion to evaluate predictive performance.

\begin{table}[H]
\centering
\caption{Out-of-sample results: AUC.}
\label{tab:oos-results}
\begin{tabular}{llllll}
\hline
                  & $h=1$   & $h=3$   & $h=6$   & $h=9$   & $h=12$  \\ \hline
Binary FAR        & 0.982 & 0.981 & 0.887 & 0.908 & 0.901 \\
Probit regression & 0.898 & 0.929 & 0.834 & 0.778 & 0.787 \\ \hline
\end{tabular}
\end{table}

\begin{figure}[H]
    \centering
    \begin{subcaptionbox}{Binary FAR\label{fig:oos-roc-bfar}}[0.45\linewidth]
        {\includegraphics[width=\linewidth]{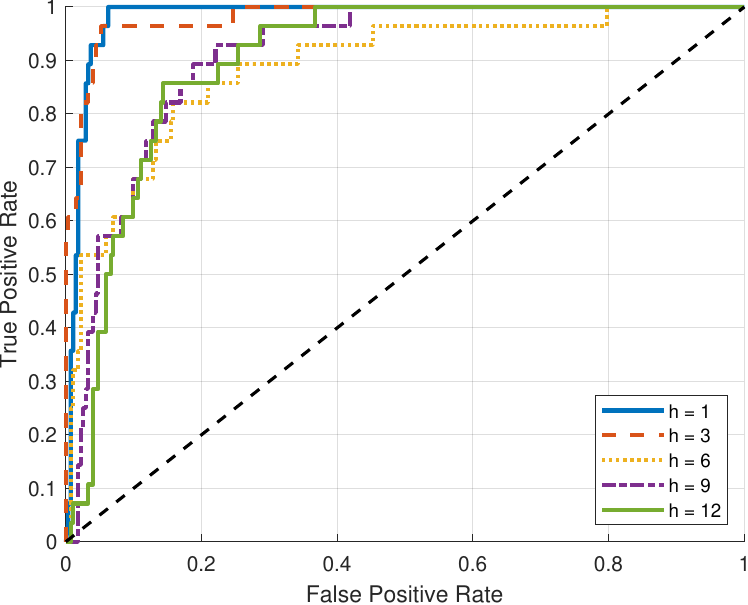}}
    \end{subcaptionbox}
    \hfill
    \begin{subcaptionbox}{Probit regression\label{fig:oos-roc-prob}}[0.45\linewidth]
        {\includegraphics[width=\linewidth]{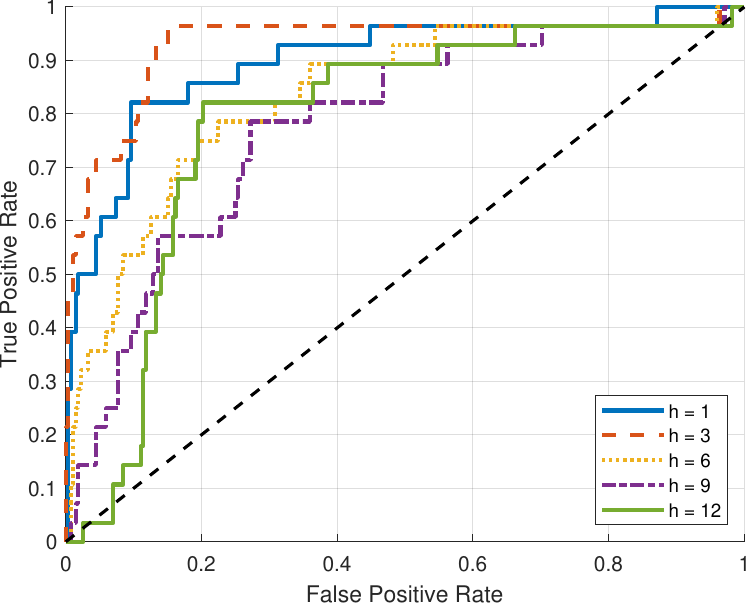}}
    \end{subcaptionbox}
    \caption{Out-of-sample ROC curves. This figure plots out-of-sample ROC curves for binary FAR and Probit regression. The 45° line represents a coin-toss classifier.}
    \label{fig:oos-roc}
\end{figure}

Table \ref{tab:oos-results} reports the out-of-sample area under the ROC curve (AUC) for the binary FAR model and the traditional Probit regression across multiple forecast horizons ($ h = 1, 3, 6, 9, 12 $). The binary FAR model consistently outperforms the Probit regression, achieving AUC values above 0.88 at all horizons and near-perfect classshowsification performance at short-term horizons (0.982 at $ h=1 $ and 0.981 at $ h=3 $). In contrast, the Probit regression shows reasonable short-term predictive accuracy but suffers a substantial decline in performance as the forecast horizon increases, with AUC values falling to 0.778 at $ h=9$ and 0.787 at $h=12$. 

Figures \ref{fig:oos-roc} display the out-of-sample ROC curves for the binary FAR and Probit models across multiple horizons, confirming the findings from Table \ref{tab:oos-results}. The binary FAR model consistently shows strong predictive accuracy, with curves for $h=1$ and $h=3$ closely approaching the top-left corner, and solid performance maintained even at longer horizons. In contrast, the Probit model exhibits weaker classification ability, particularly at $h=9$ and $h=12$, where the curves flatten toward the diagonal. These findings underscore the forecasting advantage of incorporating high-dimensional latent information through the binary FAR framework. By exploiting a richer information set embedded in the latent factors, the binary FAR model captures more persistent and informative signals related to future business cycle phases, particularly at medium- to long-term horizons.

\begin{figure}[H]
    \centering
    \begin{subfigure}[b]{\textwidth}
    \centering
    \includegraphics[width=0.18\linewidth]{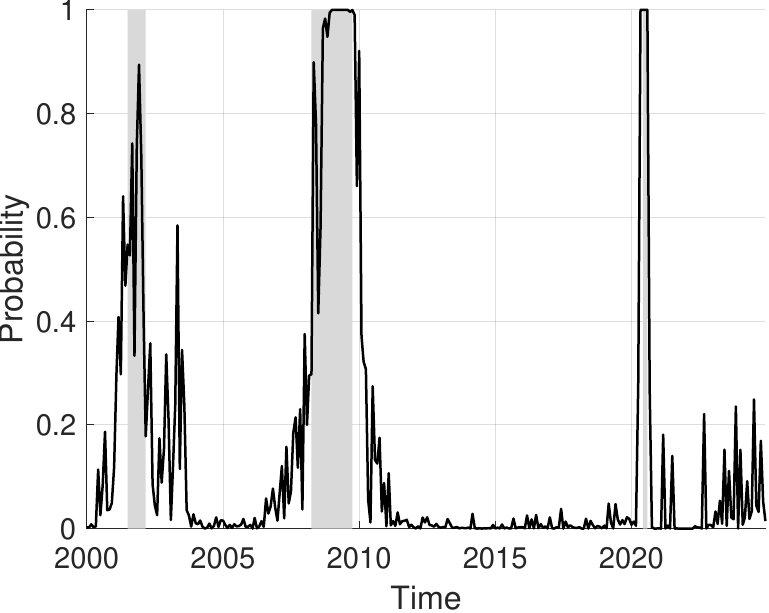}
    \includegraphics[width=0.18\linewidth]{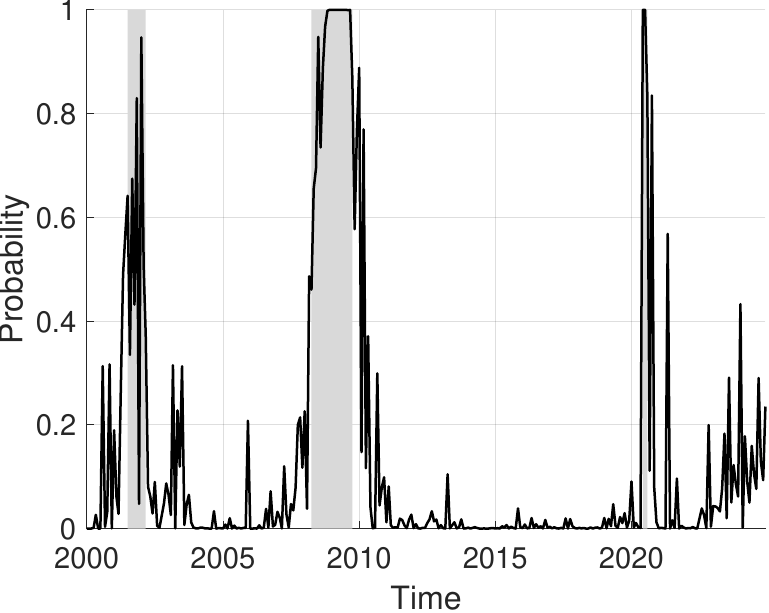}
    \includegraphics[width=0.18\linewidth]{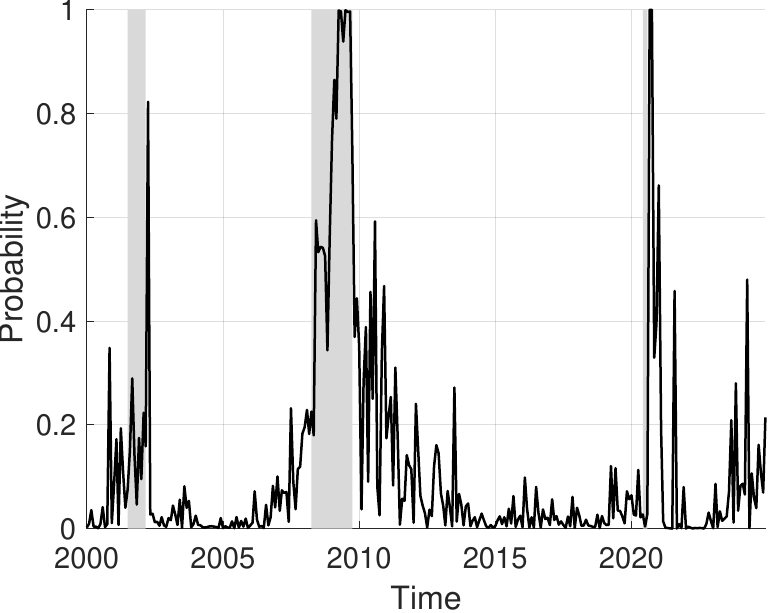}
    \includegraphics[width=0.18\linewidth]{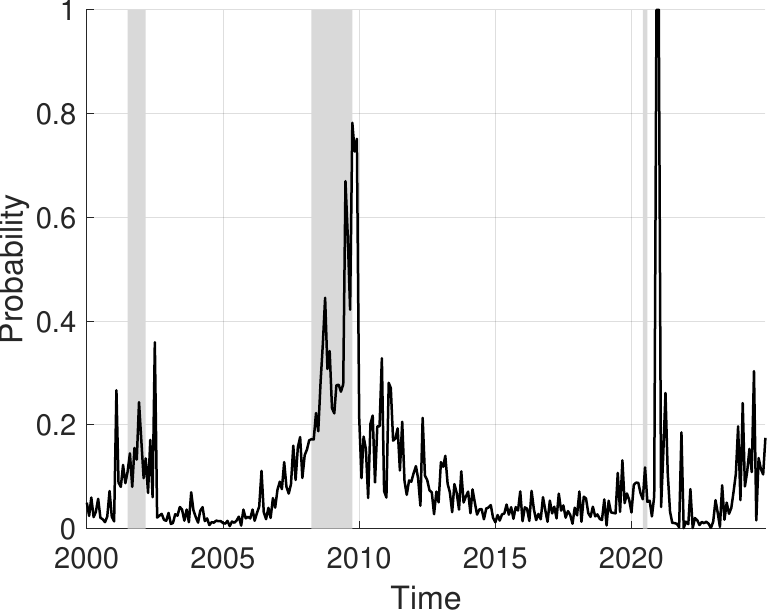}
    \includegraphics[width=0.18\linewidth]{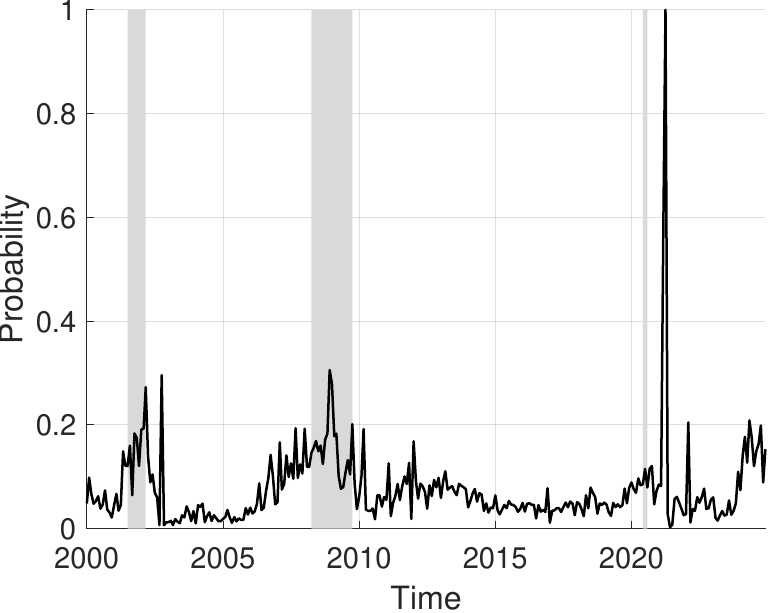}
    \caption{Binary FAR.} 
    \end{subfigure}
    
     \begin{subfigure}[b]{\textwidth}
        \centering
        \begin{minipage}[b]{0.18\linewidth}
            \centering
            \includegraphics[width=\linewidth]{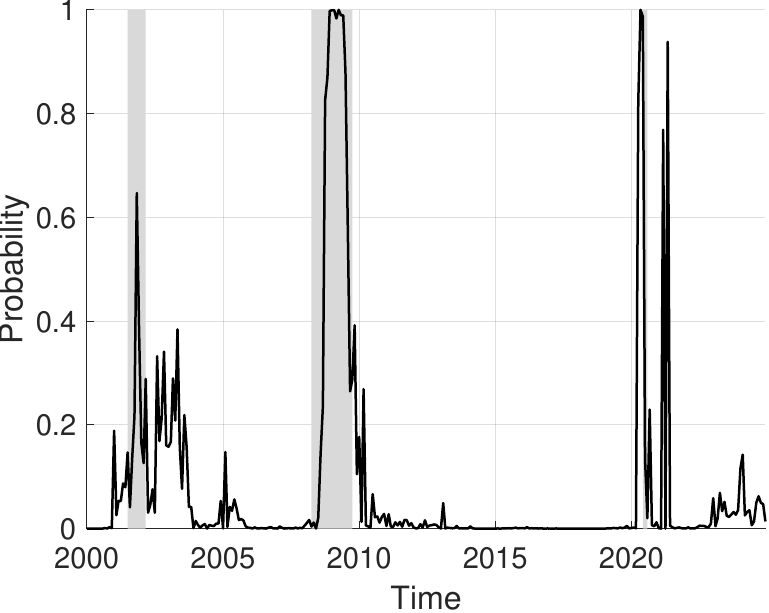}
            \small $h=1$
        \end{minipage}
        \begin{minipage}[b]{0.18\linewidth}
            \centering
            \includegraphics[width=\linewidth]{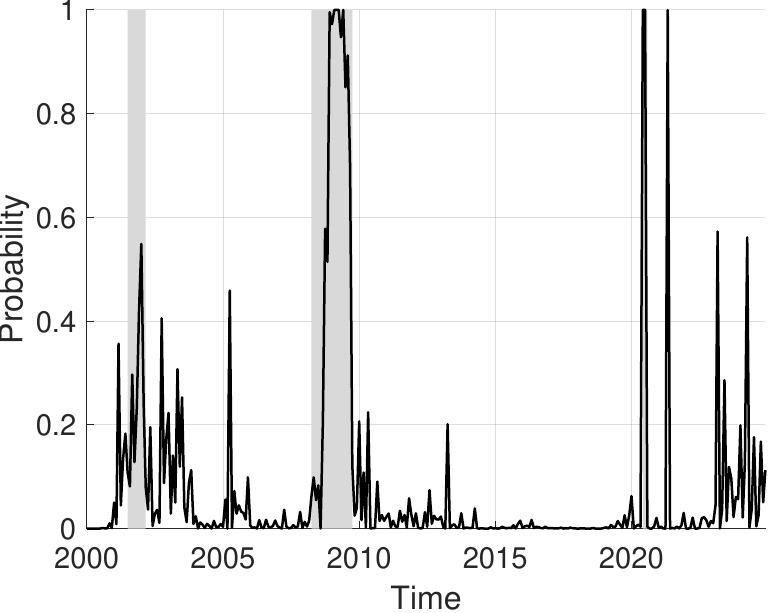}
            \small $h=3$
        \end{minipage}
        \begin{minipage}[b]{0.18\linewidth}
            \centering
            \includegraphics[width=\linewidth]{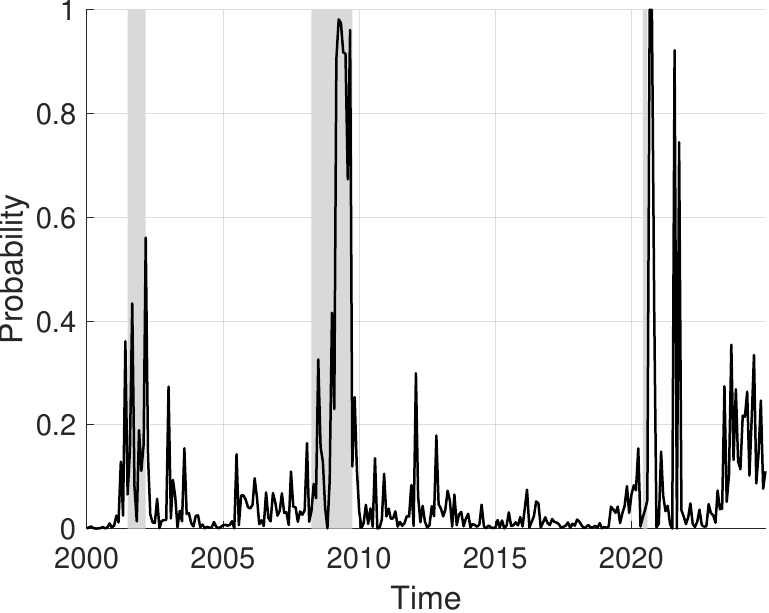}
            \small $h=6$
        \end{minipage}
        \begin{minipage}[b]{0.18\linewidth}
            \centering
            \includegraphics[width=\linewidth]{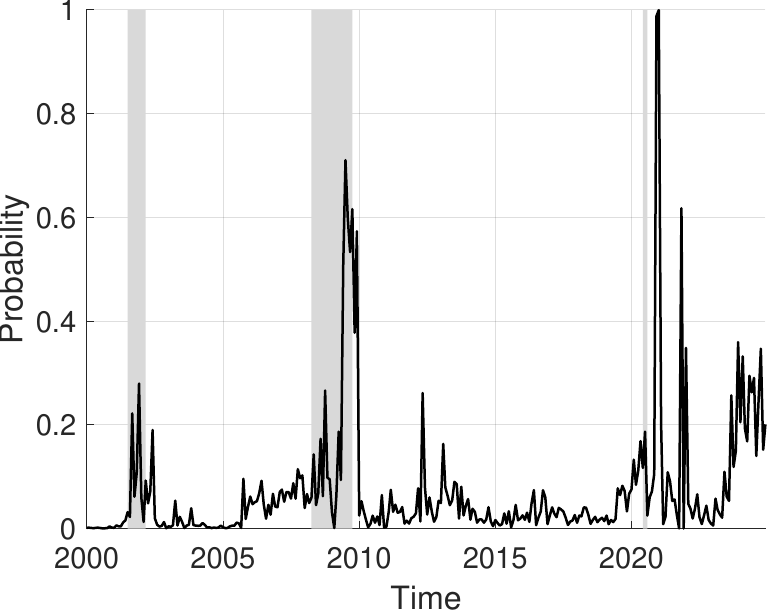}
            \small $h=9$
        \end{minipage}
        \begin{minipage}[b]{0.18\linewidth}
            \centering
            \includegraphics[width=\linewidth]{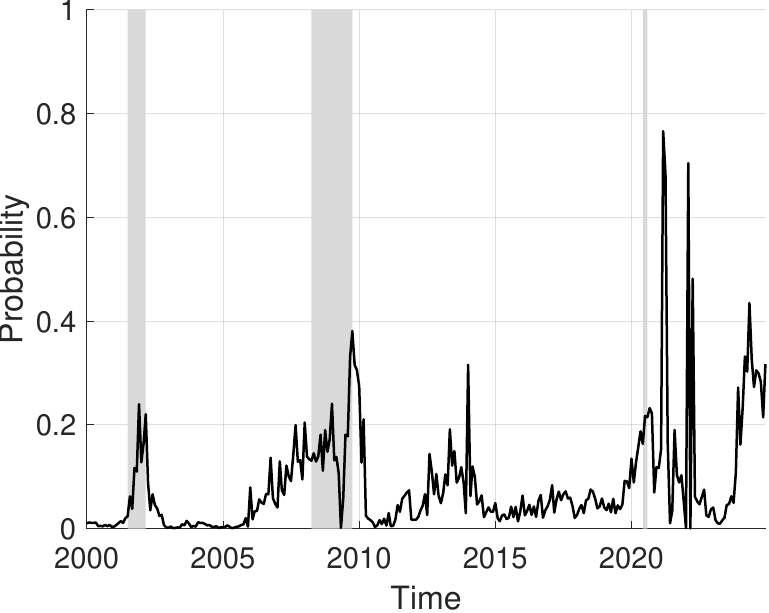}
            \small $h=12$
        \end{minipage}
        \caption{Probit regression.} 
    \end{subfigure}

    \caption{Out-of-sample forecasts. This figure plots out-of-sample forecasts of recession probabilities. NBER defined recession dates are in grey shading.}
    \label{fig:oos-forecast}
\end{figure}

Figure \ref{fig:oos-forecast} shows the out-of-sample recession probability forecasts for binary FAR and Probit regression with NBER defined recession dates plotted in grey shading. At shorter horizons ($h = 1, 3$), the binary FAR model effectively forecasts both the Dot-com crisis in 2001 and the COVID-19 pandemic in 2020. In contrast, the Probit regression fails to accurately capture the onset of the Dot-com crisis and exhibits two distinct probability peaks around 2020, resulting in a fragmented signal for the COVID-19 recession. At the longer horizon ($h = 12$), the binary FAR model continues to identify the COVID-19 episode, though with a lag of one to two months, whereas the Probit model produces a more diffuse and ambiguous forecast for the same period.

In summary, the out-of-sample results reinforce the superiority of the binary FAR model over the traditional Probit regression. By leveraging high-dimensional latent factors, binary FAR incorporates a richer and more comprehensive information set than the Probit model, which relies solely on a limited set of observable predictors. This informational advantage translates into consistently higher predictive accuracy across all horizons. Notably, the performance gap between the two models is even more pronounced in the out-of-sample setting than in-sample, highlighting the robustness and generalizability of the binary FAR model in capturing business cycle dynamics beyond the estimation sample.

\newpage
\section{Conclusion}\label{conclusion}
\noindent
\par
In this paper, we have introduced a factor-augmented model with a binary response variable by allowing for both the cross-sectional dimension and the time dimension to diverge. We have developed an MLE-based procedure to estimate regression parameters and established the corresponding asymptotic theories. From a practical perspective, the framework can be applied to conduct bank failure prediction, credit default prediction, etc. 
Following \citet{estrella1998predicting,kauppi2008predicting,christiansen2014forecasting}, in the empirical study, we focus on forecasting U.S. recessions. The proposed binary FAR model consistently outperforms conventional Probit regression across both in-sample and out-of-sample exercises, demonstrating the practical relevance and applicability of the proposed model and estimation method.
Additionally, we conduct intensive simulation studies to examine the finite sample performance of the newly proposed model and methodology.

For future work, one may consider an extension to the binary response factor-augmented regression model with functional coefficients, time-varying parameters, structure breaks, or threshold effects \citep[e.g., ][]{gu2021autoencoder,yancheng2022,chenhong2024} to capture instable relationships between the predictors and outcome variables.

\newpage
\appendix 
\setcounter{section}{0} 
\renewcommand{\theequation}{A.\arabic{equation}}
\setcounter{equation}{0}
\section{Appendix}\label{appendix}

\subsection{Lemmas and proofs}

In this appendix, we present several technical lemmas that support the proofs of the main theorems. Specifically, Lemma \ref{lemma_2.1} establishes convergence rates for the summation of error terms under the $\alpha$-mixing and weak cross-sectional dependence conditions stated in \hyperref[assumption_1]{Assumption 1}. Lemma \ref{lemma_2.2} derives asymptotic expansions for the PCA-based factor estimators, extending the theoretical results in Theorem 1 of \cite{baing2002}. Lemma \ref{lemma_2.3} further investigates the asymptotic properties of these factor estimators. Lemma \ref{lemma_2.4} and Lemma \ref{lemma_2.5} establish the convergence rates of the summation of error terms and of the first- and second-order derivatives of the individual log-likelihood functions, which are essential for deriving the convergence properties of the MLE.

\begin{lemma}\label{lemma_2.1} Let  \hyperref[assumption_1]{Assumption 1} hold. As $N,T\rightarrow\infty$,
\begin{itemize}
    \item[(a)] $\frac{1}{NT} \sum_{i,j,t,s=1} |\sigma_{ij,ts}| \leq M$ and $\max_{s,t}\frac{1}{N}\sum_{i,j=1}^{N}|\sigma_{ij,st}|\leq M$;
    \item[(b)] for each $t$, $\sum_{s=1}^{T}\vert\gamma_{ts}\vert\leq M$;
    \item[(c)] $\mathbb{E}[\frac{1}{N}\sum_{i=1}^{N}\Vert\frac{1}{\sqrt{T}}\sum_{t=1}^{T}f_te_{it}\Vert^2]\leq M$;
    \item[(d)] $ \sum_{i=1}^{N} \left( e_{is} e_{it} - \mathbb{E}(e_{is} e_{it}) \right) =O_P(\sqrt{N})$;
\end{itemize}
 where $\sigma_{ij,ts} = \mathbb{E}(e_{it}e_{js})$ and $\gamma_{ts}=\frac{1}{N}\sum_{i=1}^N\sigma_{ii,ts}$.
 
\end{lemma}

 \begin{lemma}\label{lemma_2.3} Let  \hyperref[assumption_1]{Assumption 1} hold. As $N,T\rightarrow\infty$,
 \begin{equation*}\label{lemma2.3.detail}
     \tilde{f}_t-H^\top f_t=V_{NT}^{-1} \left( \frac{1}{T} \sum_{s=1}^{T} \tilde{f}_s \gamma_{ts}
+ \frac{1}{T} \sum_{s=1}^{T} \tilde{f}_s \zeta_{st} 
+ \frac{1}{T} \sum_{s=1}^{T} \tilde{f}_s \eta_{st} 
+ \frac{1}{T} \sum_{s=1}^{T} \tilde{f}_s \xi_{st} \right),
 \end{equation*}
 where $\zeta_{st} = \frac{1}{N} E_s^\top E_t- \gamma_{ts}$, $\eta_{st} = \frac{1}{N}f_s^{\top} \Lambda^\top E_t$ and $\xi_{st} = \frac{1}{N}f_t^{\top} \Lambda^\top E_s$. Furthermore, we have
\begin{itemize}
    \item[(a)]  $T^{-1} \sum_{s=1}^{T} \tilde{f}_s \gamma_{ts} = O_p\left(\frac{1}{\sqrt{T }(\sqrt{N}\wedge\sqrt{T})}\right)$;
    \item[(b)] $T^{-1} \sum_{s=1}^{T} \tilde{f}_s \zeta_{st} = O_p\left(\frac{1}{\sqrt{N }(\sqrt{N}\wedge\sqrt{T})}\right)$;
    \item[(c)] $T^{-1} \sum_{s=1}^{T} \tilde{f}_s \eta_{st} = O_p\left(\frac{1}{\sqrt{N}}\right)$;
    \item[(d)] $T^{-1} \sum_{s=1}^{T} \tilde{f}_s \xi_{st} = O_p\left(\frac{1}{\sqrt{N}(\sqrt{N}\wedge\sqrt{T})}\right)$.
\end{itemize}
\end{lemma}

\begin{lemma}\label{lemma_2.4} Let \hyperref[assumption_1]{Assumption 1} hold. As $N,T\rightarrow\infty$,
 \begin{equation*}
 \frac{1}{T}\sum_{s=1}^T(\tilde{f}_s-H^\top f_s) f_s^\top=O_P\left(\frac{1}{\sqrt{N}}\right)+o_P\left(\frac{1}{\sqrt{T}}\right).
 \end{equation*}
\end{lemma}

\begin{lemma}\label{lemma_2.5}
Let \hyperref[assumption_1]{Assumptions 1} and \hyperref[assumption_2]{2} hold. As $N,T\rightarrow\infty$,
    \begin{equation*}
    \frac{1}{NT}\sum_{t=1}^{T-h}l_t'(u_t)\Lambda^\top E_t=O_P\left(\frac{1}{\sqrt{NT}}\right)
    \end{equation*}
    where $ u_t = \beta^\top z_t $.
\end{lemma}

\begin{lemma}\label{lemma_2.6}
Let \hyperref[assumption_1]{Assumptions 1} and \hyperref[assumption_2]{2} hold. As $N,T\rightarrow\infty$,
    \begin{equation*}\label{lemma2.6.a}
   \frac{1}{NT}\sum_{t=1}^{T-h}l_t''(\dot{u_t})f_t E_t^\top \Lambda=o_P\left(\frac{1}{\sqrt{T}}\right),
    \end{equation*}
    where $\dot{u}_t=m u_t+(1-m) \widehat{u}_t$, $m\in(0,1)$.
\end{lemma}

\begin{proof}[Proof of Lemma \ref{lemma_2.1}]~\\
(a-b). For notational simplicity, we define $\ell=|t-s|$.
For any  $(i,j)$, using the Davydov's inequality \citep[see pages 18-20 of][]{Bosq1996} and \hyperref[assumption_1]{Assumption 1}, we have
\begin{align}\label{lemma_2.1.a.1}
\left|\sigma_{ij,ts}\right|=&\left|E(e_{it}\cdot e_{js})\right|\nonumber
\\
\leq& 2\cdot\frac{4+\delta}{\delta}\cdot(2\cdot\alpha_{ij}(\ell))^{\delta/(4+\delta)}\cdot(E(e_{it})^{(4+\delta)/2})^{\frac{2}{4+\delta}}\cdot(E(e_{js})^{(4+\delta)/2})^{\frac{2}{4+\delta}}\nonumber
\\
=&C_\delta\cdot\alpha_{ij}(\ell)^{\delta/(4+\delta)},
\end{align}
where $C_\delta=\frac{4+\delta}{\delta}\cdot2^{\frac{4+2\delta}{4+\delta}}\cdot(E(e_{it})^{(4+\delta)/2})^{\frac{2}{4+\delta}}\cdot(E(e_{js})^{(4+\delta)/2})^{\frac{2}{4+\delta}}<\infty$.

By (\ref{lemma_2.1.a.1}), we obtain
    \begin{align*}
        \frac{1}{NT}\sum_{i,j=1}^N\sum_{s,t=1}^T| \sigma_{ij,ts}|\leq&O(1)\cdot\frac{1}{N}\sum_{i,j=1}^N\sum_{\ell=0}^\infty \alpha_{ij}(\ell)^{\delta/(4+\delta)}\nonumber
        \\
        =&O(1).
    \end{align*}
Therefore,  the first result in Lemma \ref{lemma_2.1}.(a) holds. Additionally, by (\ref{lemma_2.1.a.1}) and the $\alpha$-mixing conditions,
\begin{eqnarray*}
    \max_{s,t}\frac{1}{N}\sum_{i,j=1}^{N}|\sigma_{ij,st}|&\leq& O(1)\cdot\frac{1}{N}\sum_{i,j=1}^N \alpha_{ij}(0)^{\delta/(4+\delta)}\nonumber
        \\
        &=&O(1),
\end{eqnarray*}
which yields the second result in Lemma \ref{lemma_2.1}.(a).
Analogous arguments can be applied to establish Lemma \ref{lemma_2.1}.(b).

\medskip
\noindent
(c). We then proceed with the proof of Lemma \ref{lemma_2.1}.(c).
    \hyperref[assumption_1]{Assumption 1} immediately yields that  
    \[
    \mathbb{E}[f_te_{it}]=0\text{ and }\mathbb{E}[\|f_te_{it}\|^{\frac{4+\delta}{2}}]<\infty.
    \]
    By Davydov's inequality, we have
    \begin{align*}
        \mathbb{E}\left[\left\Vert\frac{1}{\sqrt{T}}\sum_{t=1}^{T}f_te_{it}\right\Vert^2\right]=&\mathbb{E}\left[\frac{1}{T}\sum_{t=1}^{T}\sum_{s=1}^{T}f_t f_s^\top e_{it} e_{is}\right]\nonumber
        \\
        =&\frac{1}{T}\sum_{t=1}^{T}\sum_{s=1}^{T}\operatorname{cov}(f_te_{it},f_se_{is})\nonumber
        \\
        \leq&O(1)\cdot \sum_{\ell=0}^\infty(\alpha_i(\ell))^{\delta/(4+\delta)}\cdot(\mathbb{E}\|f_te_{it}\|^{\frac{4+\delta}{2}})^{\frac{2}{4+\delta}}\cdot(\mathbb{E}\|f_se_{is}\|^{\frac{4+\delta}{2}})^{\frac{2}{4+\delta}}\nonumber
        \\
        =&O(1).
    \end{align*}
Therefore,  Lemma \ref{lemma_2.1}.(c) holds.

\medskip
\noindent
(d). For notational simplicity, we define
    \[
    D_i = e_{is}e_{it} - \mathbb{E}\bigl(e_{is}e_{it}\bigr)\text{~and~}D_N = \frac{1}{\sqrt{N}} \sum_{i=1}^{N}\left( e_{is}e_{it} - \mathbb{E}(e_{is}e_{it}) \right),
    \]
    By definition, it is straightforward to see that $\mathbb{E}[D_i]=0$ and $\mathbb{E}[D_N]=0$.

    To prove Lemma \ref{lemma_2.1}.(d), it suffices to show
    \begin{equation}\label{eq_r3}
       \operatorname{Var}(D_N) = \frac{1}{N}\sum_{i=1}^{N}\sum_{j=1}^{N} \operatorname{Cov}(D_i,D_j)=O_P(1). 
    \end{equation}
    For $i=j$, by  \hyperref[assumption_1]{Assumption 1}, we have
    \begin{equation}\label{eq_r1}
    \operatorname{Cov}(D_i,D_i)=\mathbb{E}[D_i^2]=O(1).
    \end{equation}
For \(i \neq j\), the dependence between $D_i$, $D_j$ can be regulated via the \(\alpha\)-mixing property in \hyperref[assumption_1]{Assumption 1}.  Using arguments that are analogous to those in (\ref{lemma_2.1.a.1}), we obtain
\begin{equation}\label{eq_r2}
\sum_{i\neq j}\left|\operatorname{Cov}(D_i,D_j)\right| \le O(1) \cdot \sum_{i\neq j} \left(\alpha_{ij}(0)\right)^{\delta/(4+\delta)}=O(N).
\end{equation}
Combining (\ref{eq_r1}) and (\ref{eq_r2}) gives (\ref{eq_r3}), which leads to the desired results in   Lemma \ref{lemma_2.1}.(d).
\end{proof}

\begin{proof}[Proof of Lemma \ref{lemma_2.3}]~\\
Lemma \ref{lemma_2.3} is a direct extension of  Lemma A.2 in \cite{bai2003}, therefore its proof is omitted here.
\end{proof}

\begin{proof}[Proof of Lemma \ref{lemma_2.4}]~\\
    \indent
 Lemma \ref{lemma_2.4} is a direct extension of  Lemma B.2 in \cite{bai2003}, therefore its proof is omitted here.
\end{proof}

\begin{proof}[Proof of Lemma \ref{lemma_2.5}]~\\
    \indent
    We write
        \begin{equation*}\label{lemma2.5.2}
            \begin{aligned}
                \frac{1}{NT}\sum_{t=1}^{T-h}l_t'(u_t)\Lambda^\top E_t=\frac{1}{NT}\sum_{t=1}^{T-h}\sum_{i=1}^Nl_t'(u_t)\lambda_i e_{it}.
            \end{aligned}
        \end{equation*}
        
    By the Cauchy-Schwarz inequality,  \hyperref[assumption_1]{Assumption 1}, and Lemma \ref{lemma_2.1}, we have
        \begin{align*}
                &\mathbb{E}\left\Vert\frac{1}{NT}\sum_{t=1}^{T-h}l_t'(u_t)\Lambda^\top E_t\right\Vert^2\nonumber             
                \\
                \leq&\frac{1}{N^2T^2}\sum_{i,j=1}^N\sum_{s,t=1}^{T-h} b_U^2\cdot\mathbb{E}\left[\Vert \lambda_i\Vert\cdot\Vert \lambda_j\Vert\right]\cdot|\sigma_{ij,ts}|\nonumber
                \\
                \leq&\frac{1}{N^2T^2}\sum_{i,j=1}^N\sum_{s,t=1}^{T-h} b_U^2\cdot\sqrt{\mathbb{E}\left[\Vert \lambda_i\Vert^2\right]\cdot\mathbb{E}\left[\Vert \lambda_j\Vert^2\right]}|\sigma_{ij,ts}|\nonumber
                \\
                \leq&O(1)\cdot\frac{1}{N^2T^2}\sum_{i,j=1}^N\sum_{s,t=1}^{T-h} |\sigma_{ij,ts}|\nonumber
                \\
                =&O\left(\frac{1}{NT}\right).
        \end{align*}
        
        Therefore, Lemma \ref{lemma_2.5} holds.
    \end{proof}

\begin{proof}[Proof of Lemma \ref{lemma_2.6}]~\\
    \indent
       By the Cauchy-Schwarz inequality, Lemma \ref{lemma_2.1}, \hyperref[assumption_1]{Assumptions 1} and \hyperref[assumption_2]{2}, we have
        \begin{align*}\label{lemma2.6.1}
                &\mathbb{E}\left\Vert\frac{1}{NT}\sum_{t=1}^{T-h}\sum_{i=1}^Nl_t'(\dot{u_t})f_t\lambda_i^\top e_{it}\right\Vert^2\nonumber
                \\
                \leq&\frac{1}{N^2T^2}\sum_{t=1}^{T-h}\sum_{i=1}^N\sum_{s=1}^{T-h}\sum_{j=1}^N b_U^2\cdot\mathbb{E}\left[\Vert f_t\Vert\cdot\Vert f_s\Vert\right]\cdot\mathbb{E}\left[\Vert \lambda_i\Vert\cdot\Vert \lambda_j\Vert\right]\cdot|\sigma_{ij,ts}|\nonumber
                \\
                \leq&\frac{1}{N^2T^2}\sum_{t=1}^{T-h}\sum_{i=1}^N\sum_{s=1}^{T-h}\sum_{j=1}^N b_U^2\cdot\sqrt{\mathbb{E}\left[\Vert f_t\Vert^2\right]\cdot\mathbb{E}\left[\Vert f_s\Vert^2\right]}\cdot\sqrt{\mathbb{E}\left[\Vert \lambda_i\Vert^2\right]\cdot\mathbb{E}\left[\Vert \lambda_j\Vert^2\right]}\cdot |\sigma_{ij,ts}|\nonumber
                \\
                \leq&O(1)\cdot \frac{1}{N^2T^2}\sum_{t=1}^{T-h}\sum_{i=1}^N\sum_{s=1}^{T-h}\sum_{j=1}^N |\sigma_{ij,ts}|\nonumber\\
                =&O\left(\frac{1}{NT}\right).
        \end{align*}
      
        Therefore, Lemma \ref{lemma_2.6} holds.
    \end{proof}

\subsection{Proofs of the main theorems}

\noindent
\begin{proof}[Proof of Lemma \ref{lemma_2.2}]~

By the definition of $\tilde{z}_t$  and  $z_t^\circ$, it suffices only to investigate the order of $\frac{1}{T} \sum_{t=1}^{T} \left\| \tilde{f}_t - H^\top f_t \right\|^2$.
Noteworthily, by the construction of the PCA estimators $\tilde{f}_t$, we have the following expansion:
\begin{eqnarray*}
     \tilde{f}_t-H^\top f_t&=&V_{NT}^{-1} \left( \frac{1}{NT} \sum_{s=1}^{T} \tilde{f}_s E_s^\top E_t
+ \frac{1}{NT} \sum_{s=1}^{T} \tilde{f}_s f_s^{\top} \Lambda^\top E_t 
+ \frac{1}{NT} \sum_{s=1}^{T} \tilde{f}_sf_t^{\top} \Lambda^\top E_s \right)
\nonumber\\
&:=&J_{1t}+J_{2t}+J_{3t}.
 \end{eqnarray*}

Using arguments that are similar to those in the proof of Lemma A.1 of \cite{bai2003} and Theorem 1 of \cite{baing2002}, we can obtain
\begin{eqnarray*}
    \frac{1}{T}\sum_{t=1}^T\|J_{1t}\|^2=O_P\left(\frac{1}{N}\right), \, \frac{1}{T}\sum_{t=1}^T\|J_{2t}\|^2=O_P\left(\frac{1}{T}\right), \,  \frac{1}{T}\sum_{t=1}^T\|J_{3t}\|^2=O_P\left(\frac{1}{N}\right).
\end{eqnarray*}
Therefore, we have
\begin{equation*}
    \frac{1}{T} \sum_{t=1}^{T} \left\| \tilde{f}_t - H^\top f_t \right\|^2=O_P\left(\frac{1}{N\wedge T}\right),
\end{equation*}
which immediately establishes Lemma \ref{lemma_2.2}.
\end{proof}

\noindent
\begin{proof}[Proof of Theorem \ref{Theorem_2.1_consistency}]~\\
It follows the first-order condition of the maximum likelihood estimation that 
\begin{align}\label{thm1.1}
    0 =\frac{\partial\left[ \frac{1}{T} \sum_{t=1}^{T-h} l_t(\beta^\top \tilde{z}_t)\right]}{\partial \beta}\Big|_{\beta=\widehat{\beta}}= \frac{1}{T} \sum_{t=1}^{T-h} l_t'(\widehat{\beta}^\top \tilde{z}_t) \tilde{z}_t.
\end{align}

Applying the Taylor expansion and  the mean value theorem for vector-valued functions  \citep[see][for example]{feng2013}, we  obtain
\begin{equation}\label{thm1.2}
    \begin{aligned}
0 =\frac{1}{T}\sum_{t=1}^{T-h} l_t'(\beta^{\circ\top}\tilde{z_t}) \tilde{z_t}
+ \frac{1}{T}\sum_{t=1}^{T-h} \left[ \int_0^1 l_t''(\tilde{z_t}^\top (\beta^\circ + s (\widehat{\beta} - \beta^\circ))) ds \right] \tilde{z}_t \tilde{z}_t^\top (\widehat{\beta} - \beta^\circ).
\end{aligned}
\end{equation}

We then further expand the first term on the right-hand side of (\ref{thm1.2}) as follows:
\begin{align*}
    \frac{1}{T} \sum_{t=1}^{T-h} l_t'(\beta^{\circ\top} \tilde{z}_t)  \tilde{z}_t &= \frac{1}{T} \sum_{t=1}^{T-h} \, l_t'(\beta^{\circ\top} z_t^\circ)  z_t^\circ 
+ \frac{1}{T} \sum_{t=1}^{T-h} \left[  \, l_t'(\beta^{\circ\top} \tilde{z}_t) -  \, l_t'(\beta^{\circ\top} z_t^\circ) \right] z_t^\circ \nonumber \\
&+ \frac{1}{T} \sum_{t=1}^{T-h} \, l_t'(\beta^{\circ\top} z_t^\circ)  (\tilde{z}_t - z_t^\circ) + \frac{1}{T} \sum_{t=1}^{T-h} \left[  \, l_t'(\beta^{\circ\top} \tilde{z}_t) - \, l_t'(\beta^{\circ\top} z_t^\circ) \right] (\tilde{z}_t - z_t^\circ)\nonumber
\\
&:=C_1+C_2+C_3+C_4,
\end{align*}
where the definitions of $C_1$, $\cdots$, $C_4$ are obvious. We study these terms one by one.

For $C_1$, recall that $u_t=\beta^{\top} z_t=\beta^{\circ\top} z_t^\circ$, simple algebra yields 
\begin{eqnarray*}
     l_t'(u_t)&=&\frac{\phi_\epsilon(u_t)(y_{t+h}-\Phi_\epsilon(u_t)) }{\Phi_\epsilon(u_t)(1-\Phi_\epsilon(u_t))}.
\end{eqnarray*}
Directly applying the law of iterated expectations, we obtain 
\begin{equation}\label{eq_r6}
    \mathbb{E}\left[\frac{1}{T} \sum_{t=1}^{T-h}  l_t'(u_t)  z_t\right]=0.
\end{equation}
Additionally, it is noteworthy that $\xi_t := l_t'(u_t)z_t$ is also an $\alpha$-mixing process under  \hyperref[assumption_1]{Assumption 1}. Using arguments that are similar to those in (\ref{lemma_2.1.a.1}), we can readily obtain 
\begin{eqnarray}\label{eq_r7}
     \mathbb{E}\left[\left(\frac{1}{T} \sum_{t=1}^{T-h}  l_t'(u_t)  z_t\right)^2\right]&=& \frac{1}{T^2} \sum_{s,t=1}^{T-h}{\rm cov}(\xi_t,\xi_s)
     \nonumber\\
     &=&O\left(\frac{1}{T}\right).
\end{eqnarray}
Combining (\ref{eq_r6}) and (\ref{eq_r7}) yields
\begin{align}\label{thm1.4}
\frac{1}{T} \sum_{t=1}^{T-h}  l_t'(u_t)  z_t^\circ=O_P\left(\frac{1}{\sqrt{T}}\right).
\end{align}

For $C_2$, by \hyperref[assumption_1]{Assumptions 1} and \hyperref[assumption_2]{2}, Lemma \ref{lemma_2.2} and Taylor expansions, we have
\begin{align}\label{thm1.5}
    \left\|\frac{1}{T} \sum_{t=1}^{T-h} \left[ \, l_t'(\beta^{\circ\top} \tilde{z}_t) - \, l_t'(\beta^{\circ\top} z_t^\circ) \right] z_t^\circ\right\|&=\left\|\frac{1}{T}\sum_{t=1}^{T-h} \left[ l_t''(\ddot{u_t})(\beta^{\circ\top} \tilde{z}_t-\beta^{\circ\top} z_t^\circ) \right]z_t^\circ\right\|\nonumber
    \\
    &\leq O(1)\cdot\frac{1}{T}\sum_{t=1}^{T-h}\left\|\beta^{\circ\top}(\tilde{z}_t-z_t^\circ) z_t^\circ \right\|\nonumber
    \\
    &\leq O(1)\cdot\left( \frac{1}{T}\sum_{i=1}^T\Vert\tilde{z}_t-z_t^\circ\Vert^2 \right)^{\frac{1}{2}} \cdot \left( \frac{1}{T} \sum_{t=1}^T \left\| z_t^\circ \right\|^2 \right)^{\frac{1}{2}}\nonumber
    \\
    &=O_P\left(\frac{1}{\sqrt{N}\wedge\sqrt{T}}\right),
\end{align}
where $\ddot{u_t}$ lies between $\beta^{\circ\top} \tilde{z}_t$ and $\beta^{\circ\top} z_t^\circ$.

We now proceed with $C_3$ and $C_4$. Similarly to (\ref{thm1.5}), we obtain
\begin{align}\label{thm1.6}
 \left\|\frac{1}{T} \sum_{t=1}^{T-h}l_t'(\beta^{\circ\top} z_t^\circ)\cdot (\tilde{z}_t - z_t^\circ)\right\|
=&O_P\left(\frac{1}{\sqrt{N}\wedge\sqrt{T}}\right),
\end{align}
and
\begin{align}\label{thm1.7}
    & \left\|\frac{1}{T} \sum_{t=1}^{T-h} \left[ l_t'(\beta^{\circ\top} \tilde{z}_t) -l_t'(\beta^{\circ\top} z_t^\circ) \right] (\tilde{z}_t - z_t^\circ)\right\|\nonumber\\
    \leq& O(1) \cdot \left(\frac{1}{T} \sum_{t=1}^{T-h}\left\| \tilde{z}_t - z_t^\circ \right\|^2\right)\nonumber
    \\
    =&O_P\left(\frac{1}{N\wedge T}\right).
\end{align}
By  (\ref{thm1.4}), (\ref{thm1.5}), (\ref{thm1.6}),  and (\ref{thm1.7}), we obtain
\begin{align}\label{thm1.8pre}
\frac{1}{T}\sum_{t=1}^{T-h}l_{t}' \left( \beta^{\circ\top}\tilde{z_t} \right) \tilde{z_t}&=\frac{1}{T} \sum_{t=1}^{T-h} l_{t}' \left( u_t \right)  z_t^\circ+O_P\left(\frac{1}{\sqrt{N}\wedge\sqrt{T}}\right)
\nonumber\\
&=O_P\left(\frac{1}{\sqrt{N}\wedge\sqrt{T}}\right).
\end{align}

We now investigate the second term of (\ref{thm1.2}). Define
\begin{equation}\label{thm1.8}
    \widehat{\Omega}_z=\frac{1}{T}\sum_{t=1}^{T-h} \left[ \int_0^1 l_{t}'' \left( \tilde{z_t}^\top (\beta^\circ + s (\widehat{\beta} - \beta^\circ)) \right) ds \right] \tilde{z}_t \tilde{z}_t^\top.
\end{equation}
Since it is required in  \hyperref[assumption_2]{Assumption 2} that $0< b_L \leq -l_{t}''(x)\leq b_U$, we have
\begin{align*}
    \left|\int_0^1 l_{t}'' \left( \tilde{z_t}^\top (\beta^\circ + s (\widehat{\beta} - \beta^\circ)) \right) ds\right|\in[b_L,b_U],
\end{align*}
uniformly for all $t$.

Using similar arguments to those in (\ref{thm1.5}), we can readily show 
\begin{eqnarray}\label{covg_estimator}
     \frac{1}{T}\sum_{t=1}^{T-h}\tilde{z}_t\tilde{z}_t^\top&=&\frac{1}{T}\sum_{t=1}^{T-h}z_t^{\circ}z_t^{\circ\top}+O_P\left(\frac{1}{\sqrt{N}\wedge\sqrt{T}}\right)
     \nonumber\\
     &=&\text{diag}(I_{p+1},H^\top)\left(\frac{1}{T}\sum_{t=1}^{T-h}z_tz_t^\top\right) \text{diag}(I_{p+1},H)+O_P\left(\frac{1}{\sqrt{N}\wedge\sqrt{T}}\right)
     \nonumber\\
     &=&\text{diag}(I_{p+1},V^{-1}Q^\top\Sigma_\lambda)\Sigma_z\text{diag}(I_{p+1},\Sigma_\lambda Q V^{-1})(1+o_P(1))
     \nonumber\\
     &&+O_P\left(\frac{1}{\sqrt{N}\wedge\sqrt{T}}\right),
\end{eqnarray}
where $\Sigma_z =\text{plim}\,T^{-1}\sum_{t=1}^{T-h}z_tz_t^\top  $, $V=\text{plim}\,V_{NT}$, $Q=\text{plim}\,T^{-1}F^\top \Tilde{F}$ and $\Sigma_\lambda=\text{plim}\,N^{-1}\Lambda^\top\Lambda$.

By (\ref{thm1.8}) and (\ref{covg_estimator}),  
\begin{eqnarray*}
    |\rho_{\min}  (\widehat{\Omega}_z)|\geq b_L \rho_{\min}  (\Sigma_z)\min\{1,\sigma_{\min}^2(V^{-1}Q^\top\Sigma_\lambda) \}+o_P(1),
\end{eqnarray*}   
where $\rho_{\min}(\cdot)$ and $\sigma_{\min}(\cdot)$ denote the minimum eigenvalue and singular value of a matrix, respectively.
Together with (\ref{thm1.2}) and (\ref{thm1.8pre}), it completes the proof of \hyperref[Theorem_2.1_consistency]{Theorem 1}.
\end{proof}

\begin{proof}[Proof of Theorem \ref{Theorem_2.2_asymptotic}]~\\
    Drawing upon the first order condition of MLE in equation (\ref{thm1.1}), we can obtain the following expansion:
\begin{align*}
        0 =&\frac{1}{T} \sum_{t=1}^{T-h} l_t'(\widehat{u}_t) \tilde{z}_t\nonumber
        \\
        =&\frac{1}{T} \sum_{t=1}^{T-h} l_t'(u_t) \tilde{z}_t+\frac{1}{T} \sum_{t=1}^{T-h}l_t''(\dot{u}_t)(\widehat{u}_t-u_t)\tilde{z}_t\nonumber
        \\
        =&\frac{1}{T} \sum_{t=1}^{T-h} l_t'(u_t)  z_t^\circ+\frac{1}{T} \sum_{t=1}^{T-h}l_t'(u_t)(\tilde{z}_t-z_t^\circ)\nonumber
        \\
        &+\frac{1}{T} \sum_{t=1}^{T-h}l_t''(\dot{u}_t) z_t^\circ z_t^{\circ\top}(\widehat{\beta}-\beta^\circ)\nonumber
        \\
        &+\frac{1}{T} \sum_{t=1}^{T-h}l_t''(\dot{u}_t)z_t^\circ(\tilde{z}_t-z_t^\circ)^\top \beta^\circ\nonumber
        \\
        &+\frac{1}{T} \sum_{t=1}^{T-h}l_t''(\dot{u}_t)z_t^\circ(\tilde{z}_t-z_t^\circ)^\top(\widehat{\beta}-\beta^\circ)\nonumber
        \\
        &+\frac{1}{T} \sum_{t=1}^{T-h}l_t''(\dot{u}_t)(\tilde{z}_t-z_t^\circ)z_t^{\circ\top}(\widehat{\beta}-\beta^\circ)\nonumber
        \\
        &+\frac{1}{T} \sum_{t=1}^{T-h}l_t''(\dot{u}_t)(\tilde{z}_t-z_t^\circ)(\tilde{z}_t-z_t^\circ)^\top \beta^\circ\nonumber
        \\
        &+\frac{1}{T} \sum_{t=1}^{T-h}l_t''(\dot{u}_t)(\tilde{z}_t-z_t^\circ)(\tilde{z}_t-z_t^\circ)^\top(\widehat{\beta}-\beta^\circ)\nonumber
        \\
        :=&A_1+A_2+\dots+A_8.
\end{align*}

Note that for $A_1$,  by (\ref{eq_r6}), \hyperref[assumption_3]{Assumption 3}, we can obtain
\begin{align}\label{A1.2}
    \lim_{T\rightarrow\infty}{\rm Var}\left(\frac{1}{\sqrt{T}} \sum_{t=1}^{T-h} l_t'(u_t)  z_t\right)=&\lim_{T\rightarrow\infty}\mathbb{E}\left[\frac{1}{T} \sum_{t=1}^{T-h}\sum_{s=1}^{T-h} l_t'(u_t)l_s'(u_s)  z_t z_s^{\top}\right]\nonumber
    =\Omega_\beta.
\end{align}

Using the $\alpha$-mixing conditions in \hyperref[assumption_1]{Assumption 1} and applying arguments that are similar to those in the proof of Lemma B.4 of \cite{gao2023}, we obtain
\begin{equation}\label{ResultA1}
    \sqrt{T}A_1=\frac{1}{\sqrt{T}} \sum_{t=1}^{T-h} l_t'(u_t)  z_t\overset{D}{\rightarrow} \mathcal{N}(0,\Omega_\beta ).
\end{equation}

We now show that $A_2$ is asymptotically negligible.
Expanding this term, we write
\begin{align*}
A_2=&\frac{1}{T} \sum_{t=1}^{T-h}l_t'(u_t)(\tilde{z}_t-z_t^\circ)\nonumber
\\
=&\frac{1}{T} \sum_{t=1}^{T-h}l_t'(u_t)[\textbf{0}_{p+1}^\top,(\tilde{f}_t-f_t^\circ)^\top]^\top.
\end{align*}
Directly applying Lemma \ref{lemma_2.3}, we obtain
\begin{align*}
\frac{1}{T} \sum_{t=1}^{T-h}l_t'(u_t)(\tilde{f}_t-f_t^\circ)
=&\frac{1}{T} \sum_{t=1}^{T-h}l_t'(u_t)(\tilde{f}_t-H^\top f_t)\nonumber
            \\
=&V_{NT}^{-1}\cdot \frac{1}{T} \sum_{t=1}^{T-h}l_t'(u_t)(B_{1t}+B_{2t})+o_P\left(\frac{1}{\sqrt{T}}\right),
\end{align*}
where we define $B_{1t}=\frac{1}{T}\sum_{s=1}^{T}(\tilde{f}_s-H^\top f_s)\eta_{st}$ and $B_{2t}= \frac{1}{T}\sum_{s=1}^{T}H^\top f_s\eta_{st}$.

    For the term with $B_{1t}$, by Lemma \ref{lemma_2.5} and Lemma \ref{lemma_2.6}, we have
\begin{align}\label{A2.2}
\frac{1}{T} \sum_{t=1}^{T-h}l_t'(u_t) B_{1t}
=&\left(\frac{1}{NT}\sum_{t=1}^{T-h}l_t'(u_t)\Lambda^\top E_t\right)\cdot\left(\frac{1}{T}\sum_{s=1}^{T}(\tilde{f}_s-H^\top f_s)f_s^\top\right)\nonumber\\
 =&o_P\left(\frac{1}{\sqrt{T}}\right).
\end{align}
For the term with $B_{2t}$,  we can use analogous arguments to show that
    \begin{align}\label{A2.3}
            &\frac{1}{T} \sum_{t=1}^{T-h}l_t'(u_t) B_{2t}=o_P\left(\frac{1}{\sqrt{T}}\right).
    \end{align}
By (\ref{A2.2}) and (\ref{A2.3}), 
    \begin{align*}
        A_2=o_P\left(\frac{1}{\sqrt{T}}\right).
    \end{align*}

We now expand \(A_3\) as follows:
    \begin{align}\label{A3.1}
        A_3=&\frac{1}{T} \sum_{t=1}^{T-h}l_t''(\dot{u}_t) z_t^\circ z_t^{\circ\top}(\widehat{\beta}-\beta^\circ)\nonumber
        \\
        =&\frac{1}{T} \sum_{t=1}^{T-h}l_t''(u_t) z_t^\circ z_t^{\circ\top}(\widehat{\beta}-\beta^\circ)+\left[\frac{1}{T} \sum_{t=1}^{T-h}(l_t''(\dot{u}_t)-l_t''(u_t)) z_t^\circ z_t^{\circ\top}\right](\widehat{\beta}-\beta^\circ)\nonumber
        \\
        :=&A_{3,1}(\widehat{\beta}-\beta^\circ)+A_{3,2}(\widehat{\beta}-\beta^\circ),
    \end{align}
where the definitions of $A_{3,1}$ and $A_{3,2}$ are obvious.

For $A_{3,1}$, simple algebra yields
\begin{align}\label{A3.2}
    A_{3,1}=&\text{diag}(I_{p+1},H^\top)\frac{1}{T} \sum_{t=1}^{T-h}l_t''(u_t) z_t z_t^\top\text{diag}(I_{p+1},H)\nonumber
    \\
    =&(H_0^{-1})^\top\Sigma_\beta(H_0^{-1})+o_P(1),
\end{align}
where  $H_0=\text{diag}(I_{p+1},VQ^{-1}\Sigma_\lambda^{-1})$.

For $A_{3,2}$ on the right-hand side of  (\ref{A3.1}), by  Taylor expansion and Cauchy–Schwarz inequality, we have
\begin{align}\label{A3.3}
        |A_{3,2}|=&\frac{1}{T}\left\| \sum_{t=1}^{T-h}(l_t''(\dot{u}_t)-l_t''(u_t)) z_t^\circ z_t^{\circ\top}\right\|\nonumber
        \\
        =&\frac{1}{T} \left\|\sum_{t=1}^{T-h}l_t'''(\ddot{u}_t)(\dot{u}_t-u_t) z_t^\circ z_t^{\circ\top}\right\|
        \nonumber\\
        \leq&\frac{1}{T}\left(\sum_{t=1}^{T-h}|\dot{u}_t-u_t|^2  \right)^{\frac{1}{2}}\left(\sum_{t=1}^{T-h}\left\|l_t'''(\ddot{u}_t)z_t^\circ z_t^{\circ\top}\right\|^2\right)^{\frac{1}{2}}
        \nonumber\\
        =&o_P(1),
\end{align}
where $\ddot{u}_t$ lies between $\dot{u}_t$ and $u_t$, and the last equality holds by Theorem \ref{Theorem_2.1_consistency} and Lemma \ref{lemma_2.3}.

Combining (\ref{A3.2}) and (\ref{A3.3}) gives
\begin{equation*}
    A_3=(H_0^{-1})^\top\Sigma_\beta(H_0^{-1})(\widehat{\beta}-\beta^\circ)+o_P\left(\left\|\widehat{\beta}-\beta^\circ\right\|\right).
\end{equation*}

For notational simplicity, we define $W_t=(1,w_t^\top)^\top$. With this notation, we can further write $A_4$ as
    \begin{align*}
            A_4=&\frac{1}{T} \sum_{t=1}^{T-h}l_t''(\dot{u}_t)z_t^\circ(\tilde{z}_t-z_t^\circ)^\top \beta^\circ\nonumber
            \\
            =&\frac{1}{T} \sum_{t=1}^{T-h}l_t''(\dot{u}_t)\left( \begin{smallmatrix}  W_t\\ H^T f_t \end{smallmatrix} \right)(\textbf{0}_{p+1}^\top,(\tilde{f}_s-H^\top f_s)^\top)  \beta^\circ\nonumber
            \\
            =&\frac{1}{T} \sum_{t=1}^{T-h}l_t''(\dot{u}_t)\left( 
            \begin{array}{c:c}
            \textbf{0}_{(p+1)\times (p+1)}& W_t  (\tilde{f}_t - H^\top f_t)^\top\nonumber \\ 
            \hdashline
            \textbf{0}_{d\times(p+1)} & H^\top f_t  (\tilde{f}_t - H^\top f_t)^\top
            \end{array} 
            \right) \beta^\circ\nonumber
            \\
            =&\left( 
\begin{array}{c:c}
\textbf{0}_{(p+1)\times (p+1)} & \frac{1}{T} \sum_{t=1}^{T-h} l_t''(\dot{u}_t)  W_t  (\tilde{f}_t - H^\top f_t)^\top \\ 
\hdashline
\textbf{0}_{d\times(p+1)} & \frac{1}{T} \sum_{t=1}^{T-h} l_t''(\dot{u}_t)  H^\top f_t  (\tilde{f}_t - H^\top f_t)^\top
\end{array} 
\right)  \beta^\circ.
\end{align*}   
For the lower-right block matrix, by the expansion that is derived in Lemma \ref{lemma_2.3},
    \begin{align*}
            \frac{1}{T} \sum_{t=1}^{T-h} l_t''(\dot{u}_t)  H^\top f_t  (\tilde{f}_t - H^\top f_t)^\top
            =\frac{1}{T} \sum_{t=1}^{T-h} l_t''(\dot{u}_t) H^\top f_t\left(B_{1t}^\top+B_{2t}^\top\right) V_{NT}^{-1}+o_P\left(\frac{1}{\sqrt{T}}\right)\nonumber.
    \end{align*} 
    For the term with $B_{1t}$, by Lemma \ref{lemma_2.4} and Lemma \ref{lemma_2.6}, we have
    \begin{align}\label{A4.1}
            \frac{1}{T} \sum_{t=1}^{T-h} l_t''(\dot{u}_t) f_t B_{1t}^\top
            =&\left[\frac{1}{NT}\sum_{t=1}^{T-h} l_t''(\dot{u}_t) f_t E_t^\top \Lambda\right]\left[ \frac{1}{T}\sum_{s=1}^{T}f_s(\tilde{f}_s-H^\top f_s)^\top\right]\nonumber
            \\
            =&o_P\left(\frac{1}{\sqrt{T}}\right).
    \end{align}
     Analogously, for the term with  $B_{2t}$, by Lemma \ref{lemma_2.6} and \hyperref[assumption_1]{Assumption 1}, we can readily obtain
    \begin{align}\label{A4.2}
            \frac{1}{T} \sum_{t=1}^{T-h} l_t''(\dot{u}_t) f_t B_{2t}^\top         
            =&o_P\left(\frac{1}{\sqrt{T}}\right).
    \end{align}   
    By (\ref{A4.1}) and (\ref{A4.2}), 
    \begin{equation}\label{A4.3}
         \frac{1}{T} \sum_{t=1}^{T-h} l_t''(\dot{u}_t)  H^\top f_t  (\tilde{f}_t - H^\top f_t)^\top=o_P\left(\frac{1}{\sqrt{T}}\right).
    \end{equation}
Using arguments that are closely related to the proof of (\ref{A4.3}),   we can show that the upper-right block matrix in $A_4$ is also asymptotically negligible.    
 Therefore, we establish that $A_4=o_P\left(\frac{1}{\sqrt{T}}\right).$


For $A_5$, by Lemma \ref{lemma_2.3},  the condition $\frac{\sqrt{T}}{N}\rightarrow0$, and the Cauchy–Schwarz inequality, we have
\begin{align*}
        \|A_5\|
        \leq&\frac{1}{T} \left(\sum_{t=1}^{T-h} l_t''(\dot{u}_t)^2\Vert\tilde{z}_t-z_t^\circ\Vert^2\right)^{\frac{1}{2}}\left( \sum_{t=1}^{T-h}\Vert z_t^\circ\Vert^2\right)^{\frac{1}{2}}\cdot\Vert\widehat{\beta}-\beta^\circ\Vert\nonumber
        \\
        =&o_P\left(\frac{1}{\sqrt{T}}\right).
\end{align*}
Using similar arguments, we can show that $A_6$, $A_7$ and $A_8$ are also asymptotically negligible with the probability order of $o_P\left(\frac{1}{\sqrt{T}}\right)$.

In summary of the established results, we obtain
\begin{align}\label{thm2.2}
    0=\frac{1}{T} \sum_{t=1}^{T-h} l_t'(u_t)  z_t^\circ+\frac{1}{T} \sum_{t=1}^{T-h}l_t''(u_t) z_t^\circ z_t^{\circ\top}(\widehat{\beta}-\beta^\circ)+o_P(\|\widehat{\beta}-\beta^\circ\|).
\end{align}
Reorganizing the terms in (\ref{thm2.2}),  we obtain
\begin{equation*}
    \sqrt{T}(\widehat{\beta}-\beta^\circ)=\left(\frac{1}{T} \sum_{t=1}^{T-h}l_t''(u_t) z_t^\circ z_t^{\circ\top}\right)^{-1}\left(\frac{1}{\sqrt{T}} \sum_{t=1}^{T-h} l_t'(u_t)  z_t^\circ\right)+o_P\left(1\right).
\end{equation*}
Together with (\ref{ResultA1}) and (\ref{A3.2}), it leads to the desired result in \hyperref[Theorem_2.2_asymptotic]{Theorem 2}.

\end{proof}

\begin{proof}[Proof of Theorem \ref{Theorem_2.3_asymptotic}]~\\
By Taylor expansion, it is straightforward to see 
\begin{eqnarray*}
\Phi_{\epsilon}(\widehat{u}_t)-\Phi_{\epsilon}(u_t)&=&\phi_\epsilon(\dot{u}_t)(\widehat{u}_t-u_t)
\nonumber\\
&=&\phi_\epsilon(\dot{u}_t)(\widehat{\beta}-\beta^\circ)^\top z_t^\circ+\phi_\epsilon(\dot{u}_t)(\tilde{z}_t-z_t^\circ)^\top \beta^\circ
\nonumber\\
&&+\phi_\epsilon(\dot{u}_t)(\widehat{\beta}-\beta^\circ)^\top(\tilde{z}_t-z_t^\circ)
\nonumber\\
&:=&D_{1t}+D_{2t}+D_{3t},
\end{eqnarray*}
where $\dot{u}_t$ lies between $\widehat{u}_t$ and $u_t$.

By \hyperref[assumption_2]{Assumption 2}, Lemma \ref{lemma_2.2} and Theorem \ref{Theorem_2.2_asymptotic}, we can readily obtain 
\begin{eqnarray*}
D_{1t}=O_P\left(\frac{1}{\sqrt{T}}\right),\,\, D_{2t}=O_P\left(\frac{1}{\sqrt{N}\wedge \sqrt{T}}\right),\,\,D_{2t}=O_P\left(\frac{1}{\sqrt{T}(\sqrt{N}\wedge \sqrt{T})}\right).
\end{eqnarray*}

Therefore, 
\begin{align*}
   \Phi_{\epsilon}(\widehat{\beta}^\top \tilde z_t)-\Phi_{\epsilon}(\beta^\top z_t) =O_P\left(\frac{1}{\sqrt{N}\wedge\sqrt{T}}\right),
\end{align*}
which completes the proof of Theorem \ref{Theorem_2.3_asymptotic}.

\end{proof}

\newpage
{\footnotesize
\setlength{\bibsep}{2.pt plus 0ex}
\bibliography{cite.bib}
}




\end{document}